\documentclass[aps,prb,notitlepage,superscriptaddress,reprint]{revtex4-2}

\usepackage[colorlinks=true,allcolors=blue]{hyperref}
\usepackage{siunitx}
\usepackage{graphicx}
\usepackage{amsmath}
\usepackage{bm}
\usepackage{xcolor}
\usepackage{colordvi}
\usepackage{color}

\newcommand{\bq}{\begin{equation}}
\newcommand{\ee}{\end{equation}}
\newcommand{\fr}[2]{\frac{#1}{#2}}
\newcommand{\eps}{\varepsilon}

\renewcommand{\vec}[1]{\mathbf{#1}}

\begin{document}
\title{Theory of two-electrons optics experiments with smooth
potentials: Flying electron molecules}

\author{P. G. Silvestrov}
\affiliation{Institut f\"ur Mathematische Physik, Technische Universit\"at Braunschweig, D-38106 Braunschweig, Germany}
\author{Vyacheslavs Kashcheyevs}
\affiliation{Department of Physics, University of Latvia, Riga, LV-1004, Latvia}
\author{Patrik Recher}
\affiliation{Institut f\"ur Mathematische Physik, Technische Universit\"at Braunschweig, D-38106 Braunschweig, Germany}
\affiliation{Laboratory for Emerging Nanometrology Braunschweig, D-38106 Braunschweig, Germany}

\begin{abstract}
Recent experimental progress in development of on-demand sources of electrons
propagating 
along depleted quantum Hall edge channels has enabled
creation and characterization of sufficiently compact single- and two-electron
distributions with picosecond scale control and
the possibility of measuring details of these distributions. Here,
we consider the 
effects
of the long-range Coulomb interaction between two electrons on the
real time evolution of such distributions in the experimentally
relevant case of smooth guiding and quantum point contact (QPC)
potentials. Both Hanbury Brown and Twiss (HBT) and Hong-Ou-Mandel (HOM) setups
are investigated. The theoretical consideration
takes advantage of 
the separation of degrees of freedom
leading to the independent motion of the center of mass and the relative motion. The most
prominent effect of this separation is the prediction of molecular bound
states, into which two electrons can become trapped and propagate as a pair along the center of
mass trajectory while simultaneously rotating around each other.
The existence of a
number of such
molecular bound states should naturally strongly affect the outgoing
electrons' distribution in the HBT experiment leading to bunching. But also in the HOM setup where colliding electrons are initially spatially separated, we predict new
effects due to the quantum tunneling of two electrons colliding
at the QPC into the joint molecular bound states. The lifetime of these quasi-bound states is shown to depend on the symmetry of the orbital wave function of the two-electron state giving rise to means to distinguish spin-triplets from spin-singlets (enabling the creation of electronic Einstein-Podolsky-Rosen (EPR) pairs). As a characteristic signature of the paired states we investigate the probability for both
injected electrons to stay a long time at the QPC.
\end{abstract}

\maketitle

\section{Introduction}\label{Introduction}

The field of electron quantum optics aims at developing control of quantum transport experiments using setups and approaches of photonic quantum optics \cite{Bocquillon2014,Bauerle2018}. Static electron sources connected to quantum Hall edge channels could demonstrate the Pauli exclusion principle in mesoscopic conductors \cite{Henny1999,Oliver1999}. With the advent of time-dependent control over emission events \cite{Feve2007}, mesoscopic single-electron optics experiments became available. 
Signs of the exchange statistics of two electrons can be probed via scattering at beamsplitters in Hanbury Brown \& Twiss (HBT) \cite{Feve2012exp} and Hong-Ou-Mandel (HOM) \cite{Bocquillon2013} setups, similar to photons. Building on these developments, quantum tomography
\cite{Jullien2014} of arbitrary excitations~\cite{Bisognin2019,Rouseel-2021} in a modulated Fermi
sea of non-interacting electrons has been demonstrated. However, electrons are charged particles and so interaction effects have to be considered with exchange effects on equal footing in contrast to the case of photons. Another quantum property that naturally comes along with interacting particles is entanglement. Indeed it has been shown theoretically that the Coulomb interaction between electrons that collide at a beamsplitter can be employed for either creating \cite{Oliver2002,Saraga2004,Saraga2005} or detecting entanglement \cite{Schroer2014, RyuSim2022b}. A complimentary regime for electron quantum optics  with ballistic
electrons isolated in energy, time and space from the parent 2D
Hall system  has  emerged more
recently~\cite{Leicht2011,Fletcher2012,Ubbelohde2015,Waldie2015,Kataoka2016a,Freise2019},
driven by the progress in generation and detection techniques for
metrological on-demand electron
sources~\cite{Kaestner2015,Giblin2019,Reifert2019}. These
techniques can generate particularly short in time (below $10$ ps)
and wide in energy ($\sim 1$ meV) wave packets that can be
interrogated using gate-defined partitioning barriers~
\cite{Fletcher2012,Waldie2015,Kataoka2016pss}, including
examination of partitioning of electron pairs in an HBT geometry
\cite{Ubbelohde2015} and demonstration of phase-space tomography
for individual wave packets~\cite{Fletcher2019,Locane2019}. Interaction effects between such isolated electrons have been recently scrutinized \cite{Pavlovska2022,Ubbelohde2022, Fletcher2022, RyuSim2022b,Fletcher2024} in the HOM geometry confirming effects of the 
strong Coulomb interaction.

In this work, we investigate the role of bound pair formation in the scattering at a beamsplitter in the lowest Landau level of quantized cyclotron motion. It is well known that in a strong magnetic field the repulsive Coulomb interaction allows for the formation of an anti-bound state of two electrons rotating around the center-of-mass \cite{Laughlin1983b}. This state is described by the two-electron limit of the celebrated Laughlin wave function~\cite{Laughlin1983a} and is stable in the absence of disorder. We show that such pairs can not only survive in the presence of a smooth guiding- or beamsplitter potential (in the HBT geometry) but can actually be created (in the HOM geometry) out of initially separated electron pairs. We concentrate on energies for two electrons close to the top of a beamsplitter so that we can approximate it by a quadratic saddle point potential. We show that even with the mutual long-range Coulomb interaction, the dynamics separates exactly into a center of mass motion and a relative motion. We project the dynamics onto the lowest Landau level manifold assuming that the fast cyclotron motion can be averaged over in favor of so called guiding coordinates $x_g$ and $y_g$ describing the drift motion in the smooth saddle point potential acting as a quantum point contact (QPC)~\cite{Fertig1987}. In terms of $x_g$ and $y_g$, the kinetic energy is completely quenched and only the potential energies and Coulomb interactions have to be kept,
but now with $x_g$ and $y_g$ being conjugate variables $[x_g , y_g] =i l_{\rm B}^2$, where $l_{\rm B}$ is the magnetic length. Energy conservation of the two-particle system turns into contour lines of constant potential energy for the relative- and center of mass motion in phase space. We choose the distribution of energies of the emitted two electrons such that they follow a critical trajectory reaching (classically or semi-classically) the saddle point in the long-time limit. Crucially,  for the relative motion, interactions split the single-electron saddle point into two interaction-induced ones, with a new region for bound motion in-between, see Figs.~\ref{fig_2} and \ref{fig_3}.

The general approach to the two-electron problem outlined in Sections \ref{section2}-\ref{section3} below has already been applied to the analysis of two-electron collisions \cite{Pavlovska2022,Ubbelohde2022,Fletcher2022}. In the work of Pavlovska \emph{et al.} \cite{Pavlovska2022}, we have solved the classical scattering problem in the HOM geometry and formulated analytic scaling relations for transmission thresholds.
In the work of Ubbelohde \emph{et al.\ } \cite{Ubbelohde2022}, these results were further developed into a fully analytical model for the analysis of counting statistics of scattering outcomes and estimation of key experimental parameters. The model was shown to correctly describe first and second order correlation signatures in the number of detected electrons as well as accurately predict the measured indicators of elastic energy exchange at near-coincident arrival.  An independent development in Ref.~\onlinecite{Fletcher2022}  has deployed a numerical solution of the corresponding equations of motion \cite{Pavlovska2022} to analyze another experimental implementation of on-demand electron collisions, with convincing agreement between the model and the measurements. 
In this paper, we focus on two major ideas: scattering of anti-bound two-electron states (``molecules'', as outlined above), and universal time-domain behavior in a smooth beamsplitter potential.

In the \emph{HBT geometry}, two electrons are considered to be paired into a bound state with the center of mass being that of a (rotating) molecule. In this case, we show that such molecules in the vicinity of the critical trajectory exhibit a universal long time tail $\sim e^{-\lambda t}$ for the pair to stay at the beamsplitter with $\lambda$ being the Lyapunov exponent of the saddle point potential introduced precisely later.  The decay rate of this probability only depends on the geometry of the beamsplitter potential (through $\lambda$) and is the same as it would be for a single electron and is therefore a direct test for the correlated two-particle pair to be bound into a molecule. Two independent electrons would show a more rapid decay of the probability to be simultaneously delayed at the beamsplitter as $\sim e^{-2\lambda t}$.

In contrast, in the {\it HOM geometry}, electrons are initially uncorrelated and the nearly critical trajectories are classically open visiting either one or both interacting saddle points for the relative coordinate before they split apart. That  leads to a modified long-time tail, $\sim \exp(-\lambda_It)$ or $\sim \exp(-\lambda_It/2)$, depending on whether the energy is below or above the critical level line. Here $\lambda_I$ is the Lyapunov exponent of the interaction-induced saddle point  and will be defined exactly below. Note that  $\lambda_I>\lambda$ so that the center of mass and the relative coordinate  distributions decay with different rates away from the center of the beamsplitter.  Further, we consider the conditions on the two initial energies for the two electrons to visit the interacting saddle points as a function of their delay times and discuss the mechanism of energy exchange between the two electrons as measured experimentally~\cite{Ubbelohde2022}. Quantum mechanically, the relative coordinate can tunnel into molecular quasi-bound states of different orbital symmetries and lifetimes temporally separating singlet from triplet bound states. These features allow us to identify signatures of the beamsplitter potential, interaction effects and the formation of (possibly entangled) electron pairs in the time domain of electron quantum optics experiments. Time-resolved detection of individual propagating electrons in quantum Hall edge channels has 
become possible recently \cite{Waldie2015,Kataoka2016a,Fletcher2019}.

The structure of the paper is as follows. In Section \ref{section2}, we introduce our model of two interacting electrons in the presence of a strong magnetic field and subjected to a saddle point potential. In Section \ref{section3}, we demonstrate the exact separation of the center of mass from the relative motion and switch to the relevant and canonically conjugate guiding center coordinates describing the drift motion of the two electrons in the saddle point potential. 
In Section~\ref{section5}, we discuss the effects of a quadratic saddle point potential on the energy-time and the time distributions, using a phase-space  picture that allows for the representation of quantum scattering dynamics in terms of an ensemble of classical trajectories under a Liouvillian flow. The results apply equally to scattering of a single electron  and to the dynamics of the center of mass of an electron molecule. In both cases, a universal tail $\sim e^{-\lambda t}$ of the time distribution after passage through the beamsplitter is identified. 
Section \ref{section6} is devoted to analysis of the HOM geometry leading to a number of interdependent results. In Sec.~\ref{secVA}, we discuss the role of exchange symmetry on the initial conditions and the corresponding Wigner representations of the independently evolving relative and average coordinates, and examine universality of the late-time behavior of the two-electron quantum state at the beamsplitter. Next (Section \ref{secVB}), we examine the classical dynamics of the centers of sufficiently localized distributions, focusing on two special cases that admit a simple analytical treatment and an intuitive discussion of the associated interplay of energy and time differences between the two electrons. Section \ref{sec:ModelHamiltonian} introduces a simple model Hamiltonian to describe dynamics near the two interacting-induced saddle points, and illustrates the key quantum effects of resonant formation and a slow decay of the bound-pair states at the beampsplitter. Finally, in Section  \ref{secVD} we discuss how the universal delay at the beasmplitter and the  parity selection rule for the resonant wave functions could be used to filter entangled Einstein-Podolsky-Rosen (EPR) pairs.   A short conclusion in Section~\ref{secConclusion} completes the paper.

\section{Interacting electrons in a smooth potential.}\label{section2}

Propagation of two electrons 
is described by the Hamiltonian
 \bq\label{Ham_2e}
H=h(\vec r_1) +h(\vec r_2) +\fr{e^2}{\kappa r} \ .
 \ee
Here $r=|\vec r_1-\vec r_2|$, $\kappa=4 \pi \epsilon_0 \epsilon_r$ is the dielectric constant 
and
 \bq\label{H_single}
h(\vec r)=\fr{{\cal P}_x^2}{2m} +\fr{{\cal P}_y^2}{2m} +V(x,y) \ ,
\ [{\cal P}_{x},{\cal P}_{y}] =i\hbar eB \ ,
 \ee
with $V(x,y)$ the confining potential and $m$ the effective mass. 
The magnetic field $B=B_z$ enters this formula via the commutation
relation of ${\cal P}_x$ and ${\cal P}_y$ which is independent of the specific gauge of the vector potential.
For GaAs, the effective mass $m=0.067 \, m_e$, and the relative dielectric constant $\epsilon_r=12$, where $m_e$ is the electron mass in the vacuum.

\begin{figure}
\includegraphics[width=8.2cm]{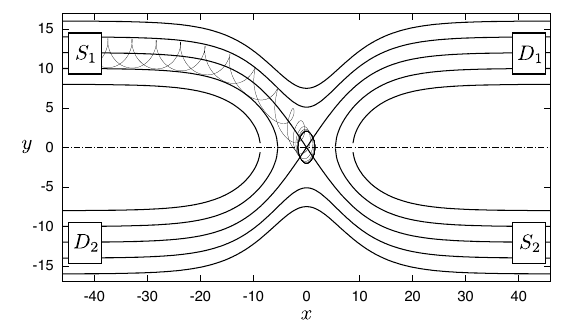} 
\caption{\label{fig_1} 
Experimental setup: HBT/HOM electron optics setup. We show the constant energy lines for the potential
$V(x,y)$ defining a wide quantum wire interrupted by a
saddle-point QPC.  
Non-interacting electrons can drift along these lines from either of the sources $S_1,S_2$ to one of the detectors $D_1,D_2$. Also shown is the
example of the classical trajectories of a pair of interacting electrons entering the
beamsplitter from the source $S_1$ and staying there forever  (conditions for this singular case are discussed in the text).
 }
\end{figure}

The smooth potential $V(x,y)$ defines the lateral confinement of electrons.
It is shown in Fig.~\ref{fig_1}, how both HBT- and HOM-electron setups may be realized in a wide wire ($V(x,y)\approx C'y^2/2 + \text{const}$) interrupted by a smooth potential barrier. 
On-demand electron sources $S_1$ and $S_2$ can inject electron wave packets from opposite sides, and detectors $D_1$ and $D_2$ are used to resolve the scattering outcome.
The center of the barrier potential contains a saddle point, where one can write~\footnote{
To characterize the curvature of the
potential both experimental~\cite{Kataoka2016a} and
theoretical~\cite{Fertig1987} papers often introduce the
harmonic-oscillator-like frequency. That means
Eq.~(\ref{saddle_point}) became
$
V(x,y)=\fr{m}{2}(\omega_y^2 y^2 -\omega_x^2 x^2)$
with obvious translation rules $A=m\omega_x^2$, $C=m\omega_y^2$.
}
 \bq\label{saddle_point}
V(x,y)\approx V_0 -\fr{Ax^2}{2} +\fr{Cy^2}{2} \ .
 \ee
For electrons with energies close to $V_0$
this potential
works as a quantum point contact or a beamsplitter. 
We will mostly use $V_0$, which corresponds to the single electron
half-transmission energy, as zero energy, i.e. $V_0\equiv 0$.

For the transverse magnetic field $B=14$ T typical in the
experiments~\cite{Kataoka2016a} the magnetic length $l_{
B}=\sqrt{\hbar/eB}\approx 6.86$ nm is naturally small compared
to the scale of variation of the confining potential $V(x,y)$. In
this case, the electron's dynamics (either classical, or quantum)
may be adequately described by the Born-Oppenheimer 
approximation separating fast (cyclotron rotation) and slow
(drifting of the cyclotron guiding centers) degrees of
freedom.  
The cyclotron motion of each electron
in the experiments with strong magnetic fields is always quantized
and to a good approximation we may assume working in the subspace
of the Hilbert space formed by the $N=0$ Landau level electrons
only. In this case the dynamics of two electrons guiding centers
is quantitatively described by the effective Hamiltonian
 \bq\label{H_eff_V}
H_{\rm eff} =\hbar\omega_L +V(x_1,y_1) +V(x_2,y_2) +\fr{e^2}{\kappa |\vec
r_1-\vec r_2|} \ ,
 \ee
where now the coordinate operators obey the commutator relations~\cite{Huckestein1995}
 \bq\label{canonical}
[x_i , y_j] =i\hbar_{\rm eff}\, \delta_{ij} =i l_{\rm B}^2 \,
\delta_{ij} \, .
 \ee
 The kinetic energy of two electrons in
the magnetic field ${\cal P}^2/(2m)$ became a trivial constant in
the effective Hamiltonian Eq.~(\ref{H_eff_V}), ${\cal
P}_1^2/(2m)={\cal P}_2^2/(2m) =\hbar\omega_L /2 =\hbar eB/(2m)$,
which accounts for the electrons' energies in the lowest Landau
level.

In a similar fashion, we may introduce the effective Hamiltonian
describing the motion of a single electron in the confining
potential $V(x,y)$
 \bq\label{H_eff_1}
H_{\rm eff}^{(1)} =\hbar\omega_L/2 +V(x,y) \ .
 \ee
Here, the guiding center coordinates $x$ and $y$ are canonically
conjugated variables as in Eq.~(\ref{canonical}), which means that the
Hamiltonian $H_{\rm eff}^{(1)}$  describes a
one-dimensional problem with {\it e.g.} the ``coordinate'' $x$ and
the ``momentum'' $y$. Classical drifting trajectories for such
a problem are the equipotential contours, $V(x,y)=\text{const}$, which we
show in Fig.~\ref{fig_1} for the wire-barrier interferometer.

The saddle point $x=0,y=0$ in Fig.~\ref{fig_1} is the 
point where the topology of the trajectories changes. The two
straight-line trajectories here are the separatrix trajectories. 
There the particles approach (leave) the unstable stationary
trajectory $(x(t),y(t))=(0,0)$ exponentially in time. These
trajectories correspond to half transmission in the quantum
mechanical case. 
They are the only classical trajectories
with indeterminate scattering outcome. 

For a smooth potential $V(x,y)$, the use of the classical
approximation is parametrically justified as long as one is not interested in any features on an area
smaller than $l_B^2$.

Classical equations of motion for the guiding centers coordinates
of two electrons follow from the effective Hamiltonian
Eqs.~(\ref{H_eff_V},\ref{canonical})
 \bq\label{eq_motion}
\dot{x}_i =\fr{l_B^2}{\hbar}\fr{\partial H_{\rm eff}}{\partial
y_i}  =\fr{1}{eB}\fr{\partial H_{\rm eff}}{\partial y_i}  \ , \
\dot{y}_i =-\fr{1}{eB}\fr{\partial H_{\rm eff}}{\partial x_i} \ .
 \ee
 
These equations may serve as a starting point for
numerical investigations of two-electron transport.
Now in Fig.~\ref{fig_1}, we
show a typical example of the 
trajectories $\vec r_1(t)$ and $\vec r_2(t)$ of two interacting electrons with close starting
points $\vec r_1(0)$ and $\vec r_2(0)$. 
Formally, drawing two trajectories on a plane does not give a true
insight into the dynamics due to the lack of information about the temporal
synchronization of $\vec r_1 (t_1)$ and $\vec r_2 (t_2)$.
Nevertheless, by looking at the trajectories in Fig.~\ref{fig_1}
we can conjecture the coherent motion of some rotating extended combined ``object''. We will proceed in the next section with explaining the symmetry leading to this object.

\section{Separation of variables}\label{section3}

Separation of the Hamiltonian into parts depending on the relative
coordinate and the center of mass position
greatly simplifies the treatment of the problem of two interacting particles. 
In this section, we show how this decomposition helps us to describe
the dynamics of two electrons in strong magnetic fields and smooth
electrostatic potentials.

Transition to the relative and center of mass coordinates and
momenta in the Hamiltonian Eqs.~(\ref{Ham_2e},\ref{H_single}) is
performed as
 \begin{align}\label{coord_trans}
\vec r&=\vec r_1-\vec r_2 \ \ , \ \
\vec R= (\vec r_1 +\vec r_2)/2 \ , \\
\vec p&= (\overrightarrow{\cal P}_1 - \overrightarrow{\cal P}_2)/2
\ \ , \ \ \vec P=\overrightarrow{\cal P}_1 + \overrightarrow{\cal
P}_2 \ . \nonumber
 \end{align}
This transformation leads to the desired decomposition of the kinetic
energy as every two components of coordinate or canonical
momentum referring to different electrons commute with each other.

Interaction between electrons is naturally only a function of the
relative coordinate $\vec r$. Even more, in many cases of interest
the potential $V(x_1,y_1)+V(x_2,y_2)$ also decouples into a sum of
parts depending only on the center of mass or the relative
coordinate.

The first of such an example is the motion in the uniform electric field,
$V(x,y)=e\vec E\cdot \vec r$ with $|\vec E|=E$, where the Hamiltonian
Eqs.~(\ref{Ham_2e},\ref{H_single}) decouples as $H\equiv H_{\rm
cm}^E +H_{\rm r}^E$ with
 \bq\label{H_cm_E}
H_{\rm cm}^E=\fr{\vec P^2}{2M} +2e \vec E\cdot\vec R \ \ , \ \
H_{\rm r}^E =\fr{\vec p^2}{2\mu} +\fr{e^2}{\kappa r} \ .
 \ee
Here $M=2m$ and $\mu =m/2$ are the total and reduced masses, respectively. The
first Hamiltonian $H_{\rm cm}^E$, allowing for a simple exact
solution, describes the drift of a composite particle in a crossed
electric and magnetic field. The drift velocity can be tuned by changing the emission energy at the source, with typical values of $v_d=E/B\approx
 (3\ldots 5)\cdot 10^4$ m/s at $B=\SI{14}{T}$ as reported in Ref.~\onlinecite{Kataoka2016a}. The corresponding electric field may be
characterized by the potential drop over the magnetic length, $l_B\, 
E\approx 4.8$ mV for $B=14$ T and $v_d=5 \cdot 10^4$ m/s. 

The second Hamiltonian in Eq.~(\ref{H_cm_E}) describes  two electrons forming a two-dimensional anti-bound state in a magnetic field~\cite{Laughlin1983b}. The presence of the Coulomb
potential splits the degenerate Landau levels $N=0,1,\cdots$ into
 \bq
{\cal E}^N_n =\hbar\omega_L (N+1/2) +\eps^N_n \ .
 \ee
Binding energies  $-\eps^N_n$ are negative due to the repulsive
interaction and decrease with increasing $n$. For a strong
quantizing magnetic field one may neglect the inter-Landau level
transitions caused by the Coulomb interaction $e^2/(\kappa r)$.
Eigenfunctions of $H_{\rm r}^E$ in this approximation coincide
with the $N$-th Landau level states with given $z$-projection of
angular momentum $m=n$ found in the symmetric gauge. For the 
$N=0$ Landau level that we are interested in,
 \bq\label{eps0n}
\eps^0_n =\fr{e^2}{\kappa l_B} \fr{\sqrt{\pi}(2n)!}{2^{2n+1}
(n!)^2} \ .
 \ee
For $B=14$ T in GaAs we thus estimate the largest energy
$\eps^0_0 =\sqrt{\pi} e^2/(2\kappa l_B)= 12.3$ meV and the binding
energies for $n\gg 1$,
 \bq
\eps^0_{n \gg 1}=\fr{e^2}{2\kappa\sqrt{n}l_B} =
\fr{6.9\,\,\rm{meV}}{\sqrt{n}} \ .
 \ee
The energy gap between Landau levels is sufficiently large, $\hbar\omega_L=24.4$ meV, and
the reduction of the Hilbert space to $N=0$ states works
reasonably well even for $n=0$ and becomes even better for large
$n$.

The decomposition of the two-electron Hamiltonian
Eq.~(\ref{H_cm_E}) is exact as long as the electric field is
uniform.
For $\vec E$ slowly changing spatially, Eq.~(\ref{H_cm_E})
approximately still holds locally. This means that two
electrons in a molecular state still drift as a whole with the center of mass retracing
the equipotential line of $V(x,y)$. Here, slowly changing obviously
means a small change experienced by a drifting molecule during one
period of rotation.

For the anti-bound states with large $n$ the period of rotation of two
electrons around their center of mass is found as
$d\eps_n/dn =\hbar/T_n$ which equals to $3.5\,\,\rm{meV}/n^{3/2}$ for $B=14$ T, leading to
$T_n=0.019\times n^{3/2}$ ps. The two electrons rotate around
their center of mass at a distance $|\vec r_1 -\vec r_2|/2=
\sqrt{n}\, l_B$ with the velocity $v_n= (2.27/n)\times 10^6$ m/s.
That means that for the edges investigated in
Ref.~\cite{Kataoka2016a}, the velocity of the mutual rotation of the
electron pair equals to the center of mass drifting velocity only
for $n\approx 50$. For molecular
states with such large $n$ deviations from the uniform electric field approximation
Eq.~(\ref{H_cm_E}) must be taken into account.

Considering nonuniform in-plane electric fields is done by
adding to the potential $V(x,y)=e\vec E\cdot \vec r$ the terms
quadratic in coordinates. Remarkably, due to the symmetry with
respect to electrons permutation the quadratic part of the full
potential $V(x_1,y_1)+V(x_2,y_2)$ still decouples exactly into a sum of quadratic parts in either relative and center of mass
coordinate.

It will be enough for us to consider the quadratic potential of
the QPC Eq.~(\ref{saddle_point}), for which one readily finds
the Hamiltonians ($H=H_{\rm cm} +H_{\rm r}$) of the center of mass
 \begin{align}\label{H_cm}
H_{\rm cm} &= \fr{\vec P^2}{2M} -AX^2 +CY^2 \ , \\
H_{\rm cm}^{\rm eff} &= \fr{\hbar\omega_L}{2} -AX_g^2 +CY_g^2 \ ,
\nonumber
 \end{align}
and the relative motion 
 \begin{align}\label{H_r}
H_{\rm r} &= \fr{\vec p^2}{2\mu} -\fr{Ax^2}{4} +\fr{Cy^2}{4}
+\fr{e^2}{\kappa r} \ , \\
H_{\rm r}^{\rm eff} &=\fr{\hbar\omega_L}{2} -\fr{Ax_g^2}{4}
+\fr{Cy_g^2}{4} +\fr{e^2}{\kappa r_g} \ .\nonumber
 \end{align}
Here, the subscript ``$g$'' labels the guiding center
coordinates. Eqs.~(\ref{H_cm},\ref{H_r}) with arbitrary
coefficients $A$ and $C$ describe the dynamics of two electrons in
the most generic quadratic potential. Only for $A,C>0$ they describe the
the saddle point potential. Another important case $A=0,C>0$ is a
wide wire having a parabolic confinement and dispersive edge states on each side with
the drift velocity varying with electron's energy. Both parameter regimes appear in the setup sketched in Fig.~\ref{fig_1}, where wide wire-like sections are interrupted by the barrier implemented as a saddle-point potential.

The decomposition $H=H_{\rm cm} +H_{\rm r}$,
Eqs.~(\ref{H_cm})  and (\ref{H_r}), is exact. In a strong quantizing
magnetic field, each of the two Hamiltonians $H_{\rm cm}$ and $H_{\rm
r}$ describes the fast  cyclotron
rotation of the corresponding coordinate
followed by a slow drift of the circular orbit's guiding center
$(X_g,Y_g)$ or $(x_g,y_g)$. Similarly to
Eqs.~(\ref{H_eff_V},\ref{canonical},\ref{H_eff_1}), we may assume
that the cyclotron  rotation is taken care of by the Landau quantization
involving the large gap $\hbar\omega_L$ and we may introduce the separate
effective Hamiltonians $H_{\rm cm}^{\rm eff}$ and $H_{\rm r}^{\rm
eff}$. This is done (as we have already shown in
Eqs.~(\ref{H_cm},\ref{H_r})) by replacing
 \bq\label{kinetic}
\vec P^2/(2M)\rightarrow \hbar\omega_L/2 \ , \ \vec
p^2/(2\mu)\rightarrow \hbar\omega_L/2 \ ,
 \ee
meaning that both fast rotational degrees of freedom are in the
ground state ($N=0$). The coordinates $X,Y,x,y$ become now the
operators of the guiding centers positions satisfying the
commutation relations~\cite{Huckestein1995}, cf.~Eq.~(\ref{canonical}),  
 \bq\label{commutation}
[X_g,Y_g] = il_B^2/2 \ , \ [x_g,y_g]= i2l_B^2 \ .
 \ee
We will consider mostly the guiding centers dynamics in the
following and omit the subscript $g$ consequently. Numerical coefficients on
the {\it r.h.s.} of the equalities Eq.~(\ref{commutation}) are
different because of the transformation to the center of mass and
relative coordinates, Eq.~(\ref{coord_trans}), not being unitary.

\begin{figure}
\includegraphics[width=7.7cm]{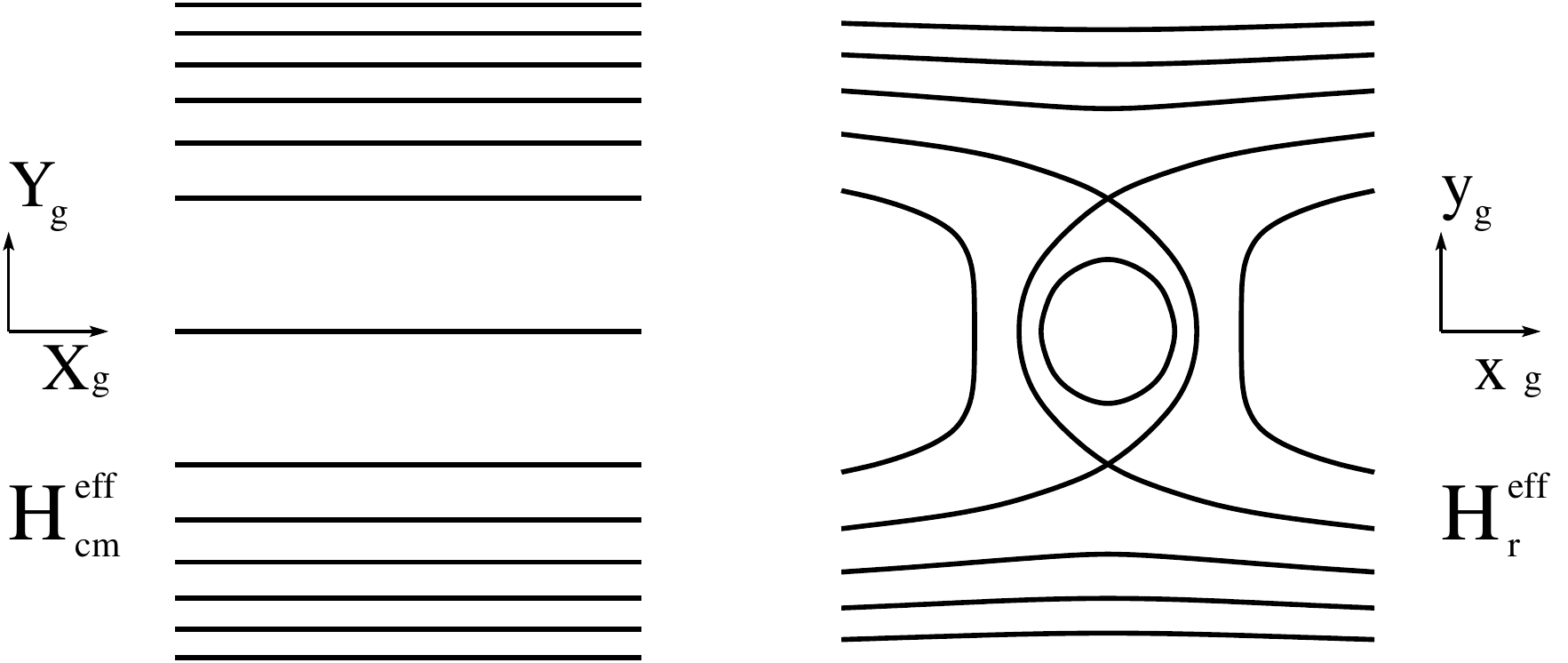} 
\caption{\label{fig_2} Contour plot for the effective
Hamiltonian $H^{\rm eff}_{\rm cm}(X_g,Y_g)$ Eq.~(\ref{H_cm}) and
$H^{\rm eff}_{\rm r}(x_g,y_g)$ Eq.~(\ref{H_r}) for $A=0$
describing the dynamics of the center of mass and
relative coordinate guiding centers of two electrons in a wide quantum wire.
 }
\end{figure}

\begin{figure}
\includegraphics[width=8.2cm]{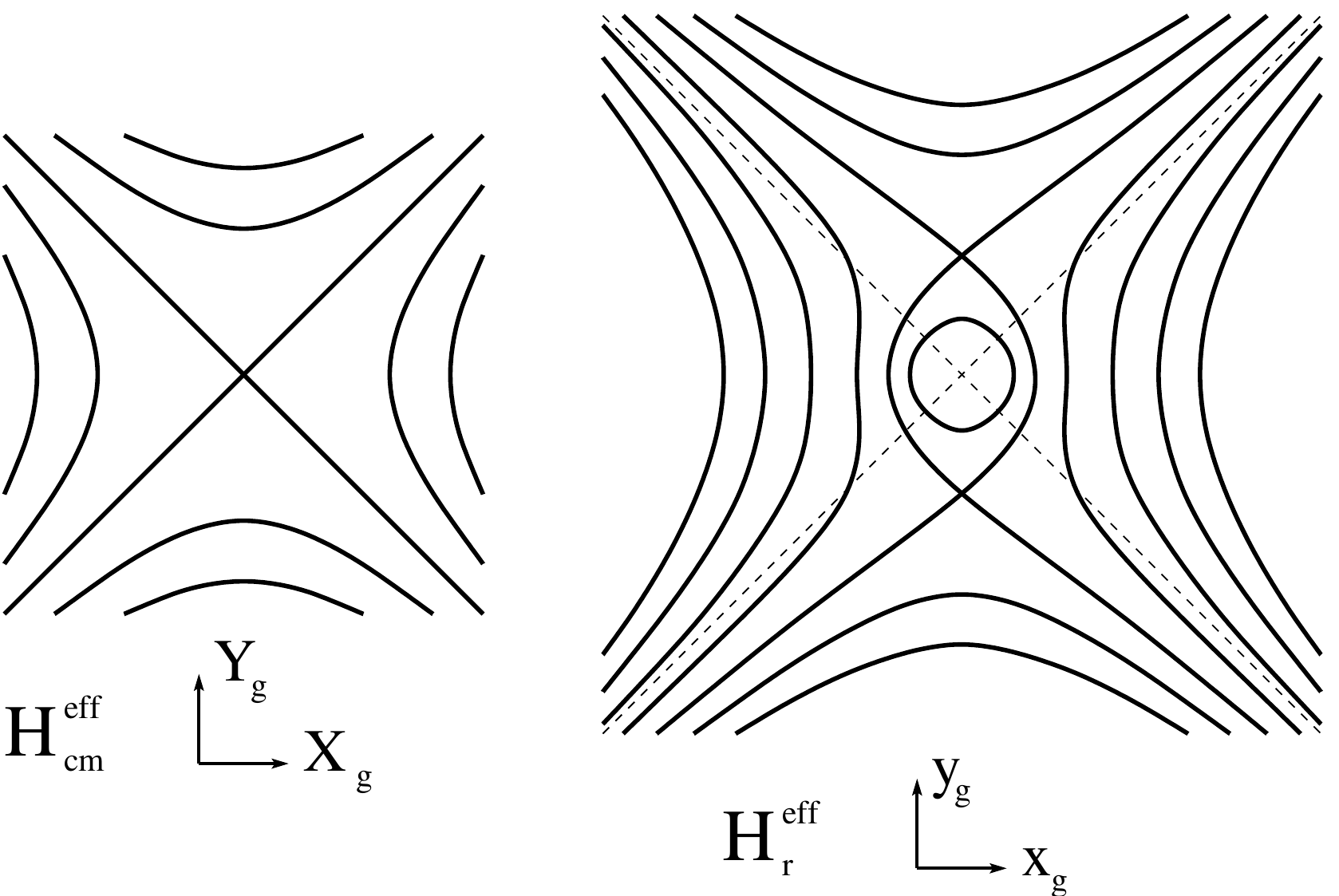} 
\caption{\label{fig_3} Contour plot for the effective
Hamiltonian $H^{\rm eff}_{\rm cm}(X_g,Y_g)$ Eq.~(\ref{H_cm}) and
$H^{\rm eff}_{\rm r}(x_g,y_g)$ Eq.~(\ref{H_r}) for $A=C$
describing the dynamics of the center of mass and
relative coordinate guiding centers of two electrons in a beamsplitter configuration.
 }
\end{figure}

Importantly, the electron-electron interaction is absent in the center of mass
Hamiltonian $H_{\rm cm}$, Eq.~(\ref{H_cm}) (as well as in $H^{\rm
E}_{\rm cm}$, Eq.~(\ref{H_cm_E})). That means that the dynamics of
the pair's center of mass coincides with the dynamics of a single
electron with the same initial conditions (think of the motion of
a pair of non-interacting electrons moving with negligible
displacement). The motion of the center of mass separates into the
fast cyclotron rotation and the slow drift of the center of this rotation along the lines of constant energy of the effective Hamiltonian $H_{\rm cm}^{\rm eff}$, Eq.~(\ref{H_cm}),
which in turn coincide with the equipotential lines of $V(x,y)$. The cyclotron 
motion of the center of mass is quantized and hence is considered in the lowest Landau level only, with constant kinetic energy
Eq.~(\ref{kinetic}). 

\subsection{Drift motion of the center of mass of two electrons in a dispersive edge}
Consider now in more details the case of the pair of electrons in
a single dispersive edge. As we discussed already, an adequate model
for such an edge is a wide wire described by
Eqs.~(\ref{H_cm},\ref{H_r}) with $A\equiv 0$ and a parabolic
transverse confinement potential $V(x,y)\sim y^2$. The topographic
map for the effective Hamiltonian for the center of mass $H_{\rm
cm}^{\rm eff}$, Eq.~(\ref{H_cm}), in this case, shown in
Fig.~\ref{fig_2} (left panel), consists of a set of straight lines $Y_g=const$.
The distance between lines is inversely proportional to $|Y_g|$ and the
drift velocity is linear in the coordinate $v_{x_{\rm cm}}=CY_g$.

The curvature $C$ of the electric potential along the depleted edge has ben estimated experimentally from time-of-flight measurements~\cite{Kataoka2016a} and from the energy dependence of LO phonon emission~\cite{Emary2019}.  Reported values are  $\hbar \omega_y = \hbar \sqrt{C/m} = \SI{1.8}{meV}$ to  $\SI{2.7}{meV}$~\cite{Kataoka2016a} and $\hbar \omega_y =\SI{2.7}{meV}$ \cite{Ubbelohde2022}, 
and may vary depending on the sample and fabrication details.

\subsection{Drift motion of the relative coordinate of two interacting electrons in a dispersive edge}
The situation of the relative coordinate Eq.~(\ref{H_r}) is more
involved.  The
guiding center drifts along the constant energy lines of the
effective Hamiltonian $H_{\rm r}^{\rm eff}(x,y)$, which now becomes
a sum of the wire potential $V(x,y)=Cy^2/4$ and the electrons'
repulsion~$e^2/( \kappa r)$.

The topographic map of the effective Hamiltonian $H_{\rm r}^{\rm
eff}(x_g,y_g)$, Eq.~(\ref{H_r}), with  $A=0$ is shown in
Fig.~\ref{fig_2} (right panel). The new feature of this map is the existence of
two interaction induced saddle points with $(x_g,y_g)=(0,\pm y_0)$
and the finite region of closed trajectories/equipotential lines
between them. The area included inside the allowed trajectories is
quantized in units of $4\pi l_B^2$ according to
Eq.~(\ref{commutation}).

For the edge with no velocity dispersion, Eq.~(\ref{H_cm_E}), the
spectrum of two electron states is discrete with the excitation 
energies given by Eq.~(\ref{eps0n}). As we see, for the dispersive
edge with the concave edge potential ($C>0$) the total number of
possible molecular bound states is finite.

From the effective Hamiltonian $H_{\rm r}^{\rm eff}(x_g,y_g)$,
Eq.~(\ref{H_r}), one easily finds the coordinate of the saddle
point $y_0 =(2e^2/(\kappa C))^{1/3}$. The total area covered by
the periodic trajectories is $A_{mol}=2 y_0^2 (3-2 \ln 2)/\sqrt{3} \approx 1.86y_0^2$. 
Parameters of the experiment analyzed in Ref.~\onlinecite{Ubbelohde2022} are $y_0=\SI{45}{nm}$ and $l_B=\SI{8}{nm}$ at $B=\SI{10}{T}$, hence we
estimate $y_0\approx 5.6 l_B$ and the total number of distinct
molecular states $N_{mol}=A_{mol}/(4\pi l_B^2)=4.6$.  The
number of molecular states is sufficiently large to justify the
use of a (semi)classical approach, but not large enough to make
quantum effects, like quantum tunneling, invisible.

In the experimental devices the curvature of the edge potential
($C$ in our notations) may vary along the edge. 
Even more,
typically in one and the same device there are two kinds of edges,
produced by etching the 2DEG and by the electron repulsion from
the metallic gates. So far, we considered the case of a concave edge
potential, $C>0$. It may easily happen (or may be designed on
purpose) that at least a part of the drifting trajectory will
traverse a region with a convex edge potential, i.e. $C<0$. All the
trajectories for the relative coordinate $\vec r(t)$ in such a region
are closed and all two-electron states states are (at least weakly) bound
molecules.

\subsection{Drift motion of the center of mass of two interacting electrons in the saddle point potential}
Our next step is to consider the electron's pair dynamics in the
saddle-point potential $V(x,y)=-Ax^2/2 +Cy^2/2$,
Eq.~(\ref{saddle_point}). Since the potential is quadratic in
coordinates, there is an exact separation of the Hamiltonian into
the center of mass $H_{\rm cm}$, Eq.~(\ref{H_cm}), and the
relative coordinate $H_{\rm r}$, Eq.~(\ref{H_r}), parts. 

The center
of mass Hamiltonian $H_{\rm cm}$ is fully quadratic in both
coordinates and momenta and as such can be solved
exactly~\cite{Fertig1987}. This solution consists of another
exact decomposition of $H_{\rm cm}$ into a sum of two quadratic
one-dimensional Hamiltonians each with its pair of canonically
conjugated variables. These pairs of variables are the
(canonically conjugated) coordinates of the cyclotron rotation of
the center of mass and (also canonically conjugated) coordinates
of the drifting center of this rotation. In the limit of a strong
quantizing magnetic field, which is our focus here, the
cyclotron rotation is projected to the $N=0$ Landau level, which is
reflected by the constant $\hbar\omega_L/2$ term in the effective
guiding center Hamiltonian $H_{\rm cm}^{\rm eff}$,
Eq.~(\ref{H_cm}). Classical trajectories for the center of mass
guiding center motion, which are the lines of constant energy
$H_{\rm cm}^{\rm eff}(X_g,Y_g)=\text{const}$ are shown in
Fig.~\ref{fig_3} (left panel). For the pure quadratic Hamiltonian these
trajectories are hyperbolas with the separatrix trajectories being straight lines. At the origin, there is a single unstable
stationary trajectory $(X_g(t),Y_g(t))\equiv (0,0)$.

Later in the paper we will often refer to the classical
trajectories for the Hamiltonians~Eqs.~(\ref{H_cm},\ref{H_r}). For
the center of mass Hamiltonian $H_{\rm cm}^{\rm eff}$ with the
commutation relations Eq.~(\ref{commutation}) one easily derives
the classical equations of motion
leading to the two-parameter family of trajectories
 \bq\label{alpha_beta}
X_g=\fr{\alpha \, e^{-\lambda t}}{\sqrt{A}} +\fr{\beta \, e^{\lambda
t}}{\sqrt{A}} \ , \ Y_g=\fr{\alpha \, e^{-\lambda t}}{\sqrt{C}}
-\fr{\beta \, e^{\lambda t}}{\sqrt{C}} \ .
 \ee
Here $\lambda =l_B^2\sqrt{AC}/\hbar =(1/eB)\sqrt{AC}$ is the
classical Lyapunov exponent. Taking $\alpha$ and
$\beta$ to be very small leads to trajectories that stay long at the unstable
stationary point. Quantitative modelling of the HOM-type experiments \cite{Pavlovska2022} yields $\lambda^{-1} \approx \SI{2}{ps}$ \cite{Fletcher2022} to $\approx \SI{4}{ps}$ \cite{Ubbelohde2022}.

\subsection{Drift motion of the relative coordinate of two interacting electrons in the saddle point potential}
The topographic map of the effective 
Hamiltonian $H^{\rm eff}_{\rm r}(x_g,y_g)$ of the relative coordinate, Eq.~(\ref{H_r}), which illustrates possible classical trajectories described
by this Hamiltonian and for a case of the beamsplitter potential
Eq.~(\ref{saddle_point})
is shown in Fig.~\ref{fig_3}b. Similar to the case of the
dispersive edge, Fig.~\ref{fig_2}b, the map shows two interaction
induced saddle points at $(x_g,y_g)=(0,\pm y_0)=(0,\pm
(2e^2/(\kappa C))^{1/3}$. Between them there is an area of
periodic classical trajectories corresponding to long-lived quantum molecular states.

Classical solutions for the relative coordinate close to one of
the interaction induced stationary points
form a two parameter family (compare to the center of mass
trajectories Eq.~(\ref{alpha_beta}))
 \bq\label{gamma_delta}
x_g=\fr{\gamma \, e^{-\lambda_{\rm I} t}}{\sqrt{A+C}} +\fr{\delta \,
e^{\lambda_{\rm I} t}}{\sqrt{A+C}} \ , \ y_g\mp y_0=\fr{\gamma \,
e^{-\lambda_{\rm I} t}}{\sqrt{3C}} -\fr{\delta \, e^{\lambda_{\rm
I} t}}{\sqrt{3C}} \ ,
 \ee
where now $\lambda_{\rm I} =(1/eB)\sqrt{3(A+C)C}$. These results
are formally valid for $|x_g|,|y_g-y_0|\ll y_0$ or for $|x_g|,|y_g+y_0|\ll
y_0$.

Since $\lambda_I >\lambda$, Eqs.~(\ref{alpha_beta},
\ref{gamma_delta}), in the vicinity of the double-stationary
unstable point (stationary for both $\vec r_g$ and $\vec R_g$) the
relative coordinates trajectories diverge faster than the center-of-mass trajectories.

\section{HBT: Dynamics of the molecule at the beamsplitter}\label{section5}

\begin{figure}
\includegraphics[width=6.2cm]{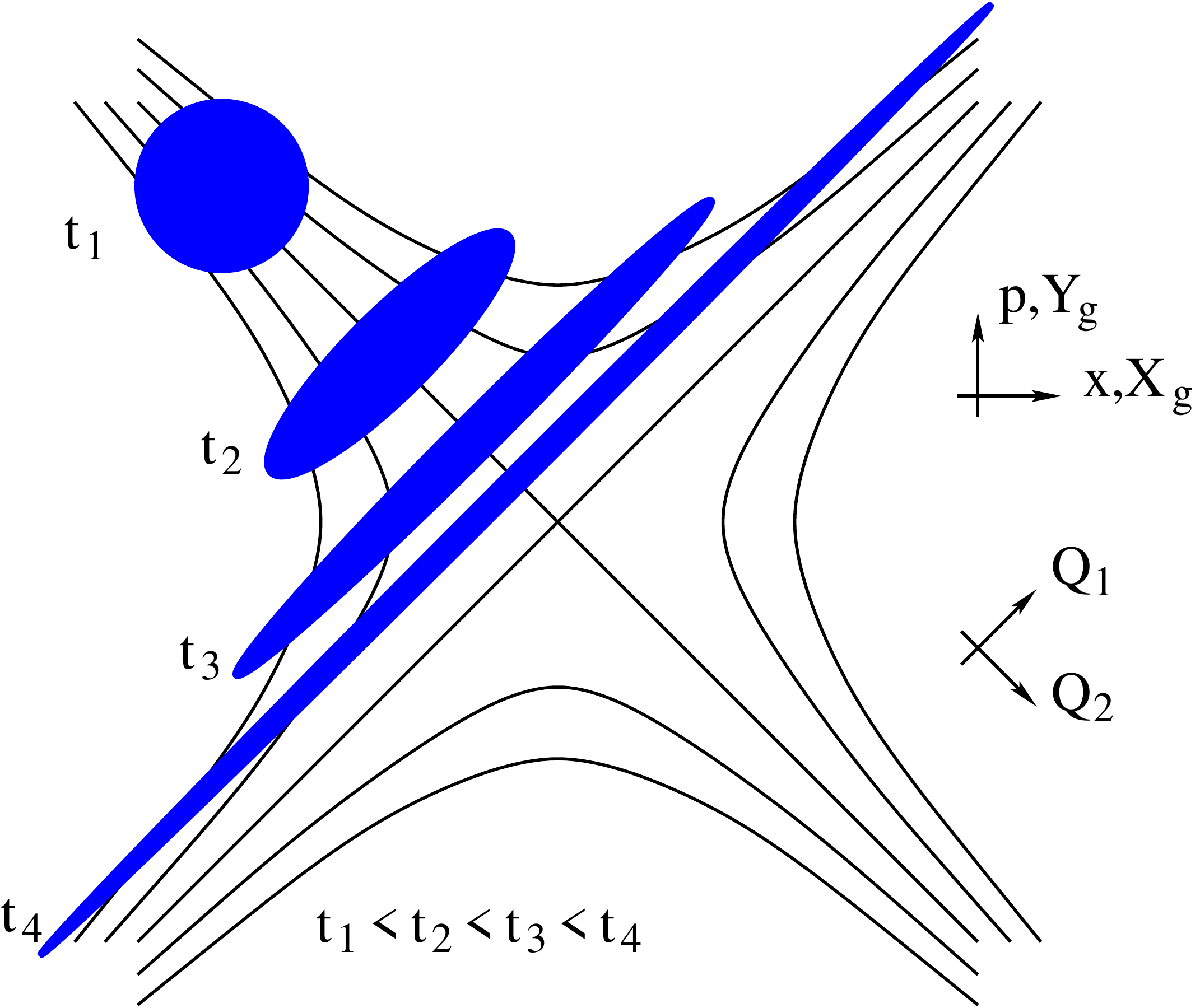} 
\caption{\label{fig_4} Time evolution of a compact distribution of a particle
on the $(x,p)$ plane in the vicinity of the parabolic potential
barrier Eq.~(\ref{H_barrier}). This may either refer to the evolution of
the classical probability density, or the density described by the
Wigner function for a quantum mechanical wave packet.
}
\end{figure}

In the previous section, we have shown that a pair of electrons in
the QH edge may form an anti-bound state and travel together as a
composite particle. Still, since in order to form a bound state
electrons should be put close to each other, we expect the quantum
wave packet describing such joint propagation to be narrow in real space, ideally
even narrow compared to the size of the beamsplitter. Such a situation can be naturally realized
in HBT geometry if the electrons are injected by the source already as pairs. 

In this section, we consider the transmission of a sufficiently
short wave packet of such composite particles (molecules) through the
beamsplitter described by the potential Eq.~(\ref{saddle_point}).
Rather counterintuitively, the quantum tunneling dynamics of the
wave packet turns out to be fully described in terms of classical
mechanics. The main result of this section is the prediction of
the universal long time probability to find the particle ({\it
i.e.} two electrons together!) inside the device $\sim e^{-\lambda
t}$, governed by the classical Lyapunov exponent $\lambda$.

The Hamiltonian describing the propagation of the composite
particle through the beamsplitter takes the generic form
 \bq\label{H_barrier}
H_{\rm barrier} = \fr{p^2}{2m} -\fr{m\lambda^2 x^2}{2} \ .
 \ee
Comparing to $H^{\rm eff}_{\rm cm}$ via Eqs. (\ref{H_cm},
\ref{commutation}) one finds the translation rules: $x=X_g$,
$p=2\hbar Y_g/l_B^2$, $[x,p]=i\hbar$,
$\lambda= l_B^2\sqrt{AC}/\hbar= (1/eB)\sqrt{AC}$,
$m=2\hbar^2/(l_B^4 C)= 2(eB)^2/C$.
In this notation, the classical Lyapunov exponent $\lambda$
appears as an imaginary frequency of the conventional harmonic
oscillator. Deviations from Eq.~(\ref{H_barrier}) in a smooth
beamsplitter potential are expected to be small until some sufficiently
large distances $x\sim L_{\rm barrier}\gg \sqrt{\hbar/m\lambda}\sim
l_B(C/A)^{1/4}$.

Scattering of a quantum particle on a potential barrier can be
fully described by the transmission $t(\eps)$ and reflection
$r(\eps)$ complex amplitudes. For a very mono-energetic wave packet with energy $\eps$ and extension long
compared to the size of the scattering region $\Delta x\sim
1/\Delta p\gg L_{\rm barrier}$, the wave packet propagates almost
without dispersion and change of shape. Its scattering properties as {\it
e.g.} transmission/reflection probabilities and delay times follow straightforwardly from
$t(\eps), r(\eps)$.   Short wave packets, $\Delta x < L_{\rm
barrier}$, are strongly deformed in the process of scattering~(two electron propagation through a  beamsplitter with interactions in the long wavepacket limit was considered in Ref.~\onlinecite{RyuSim2022b}).
Nevertheless, as we will see, for smooth potentials this
deformation is still sufficiently universal and may be described by classical dynamics without referring to quantum
scattering states. This is what we will discuss next.

\subsection{\label{sec:HBTphasespace}Phase space dynamics of a bound molecule at the beamsplitter}
Classical trajectories for the Hamiltonian Eq.~(\ref{H_barrier})
are linear combinations of exponential functions (see
Eq.~(\ref{alpha_beta})). Now we are interested not in the motion
of individual particles, but in the dynamics of an ensemble of particles.
Let $\rho(x,p,t)$ be the distribution of the ensemble. It may be a pure
classical probability distribution originated from the limited accuracy of the
initial particle parameters, or it may be a Wigner function
corresponding to a quantum wave packet,  with Weyl transform (Wigner representation) of the single-particle density matrix covering both cases  (low-purity mixed state or a pure wave function, respectively)~\cite{Locane2019}. 
The exact time evolution of this distribution in phase space formulation of quantum mechanics \cite{Hillery1984} is described by its Moyal bracket with  the Hamiltonian. For a polynomial Hamiltonian at most quadratic in $x$ and $p$ (such as 
$ H_{\rm barrier}$ above) the Moyal bracket is identical with the Poisson bracket and hence the Wigner distribution follows the Liouvillian dynamics (classical Hamiltonian flow).

The properties of this flow are most easily revealed when viewed in a different
system of coordinates determined by so called unstable and stable
directions
 \bq\label{stable_unstable}
Q_1=\sqrt{\fr{m\lambda}{2}}\, x +\fr{p}{\sqrt{2m\lambda}}  \ , \
Q_2=\sqrt{\fr{m\lambda}{2}}\, x -\fr{p}{\sqrt{2m\lambda}} \ .
 \ee
Since the time dependence of these coordinates is purely
exponential, $Q_1(t)=e^{\lambda t}Q_1(0)$, $Q_2(t)=e^{-\lambda
t}Q_2(0)$, the time dependence of the density
$\tilde\rho(Q_1,Q_2,t)= \rho(x(Q_1,Q_2),p(Q_1,Q_2),t)$ has a
simple form
 \bq\label{rho_tilde}
\tilde\rho(Q_1,Q_2,t)= \tilde\rho(Q_1 e^{-\lambda t},Q_2
e^{\lambda t},0) \ .
 \ee

\begin{figure}
\includegraphics[width=8.5cm]{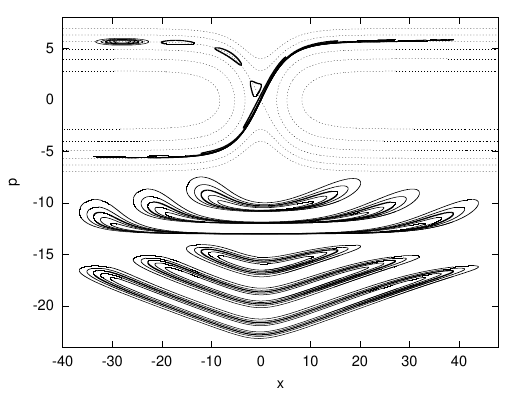} 
\caption{
\label{fig_5.3} Evolution of a classical particles distribution
upon scattering on a realistic barrier modeled by the Hamiltonian
$H_{\rm barrier}=p^2/2 +A^2/\cosh(x/A)$, $A=4$. The initial
distribution is an asymmetric Gaussian centered at the incoming
separatrix with the widths $A$ in $x$-direction and $1/A$ in
$p$-direction (this corresponds to a symmetric Gaussian
distribution in the $\eps,t$ plane). One
should remember that the classical Liouvillian evolution preserves the
local densities and areas and that only the shapes of areas
surrounded by the contour lines are deformed. {\bf Top:}
Distribution at the time slices $t=0,2,\cdots,14$ with the
topographic map of the Hamiltonian shown in the background. The
first, $t=0$, distribution is shown by the contour lines at 
multiples of $1/6$ of the maximal density. For others only the
$1/2$-maximum contour is shown. {\bf Middle:}
At every point $x$ the distribution at large times collapses to a
single line -- the outgoing separatrix $p_s(x)$, which for our
model has a form $p_s(x)=2A\sinh(x/2A)/\sqrt{\cosh(x/A)}$.
Therefore, we show for the longest times, $t=10,12,14$, the
distribution of deviations from the limiting curve multiplied by
$10$ and offset downwards, $10\times (p-p_s(x))$. The expanded vertical coordinate allows to show the full topographic map of the former Gaussian distribution
with 5 slices of the density at 1/5 of the maximum steps. {\bf Bottom:}
For the same large times probability density as a function of $\ln
(p-p_s(x))$ and $x$ showing the $e^{-\lambda t}$ tail of both
reflected and transmitted distributions (offset downwards).
 }
\end{figure}

\begin{figure}
\includegraphics[width=8.5cm]{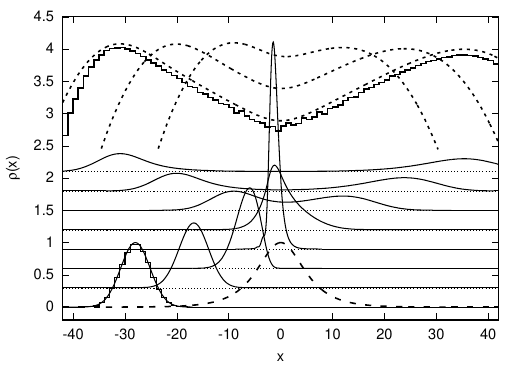} 
\caption{
\label{fig_5.2} Penetration of the short wave packet through the
realistic smooth barrier at the half-transmission energy with the same model
Hamiltonian as used for Fig.~\ref{fig_5.3}, $H_{\rm
barrier}=\hat p^2/2 +A^2/\cosh(x/A), \ \hat p=-id/dx, \ A=4$.
The potential barrier is shown explicitly by the dashed line at the
bottom (not in scale). Starting from the Gaussian wave packet
$\psi_0=\exp(-(x/A+7)^2/2+i\sqrt{2}Ax)$ we calculate the particle
density $\rho_t(x)=|\psi_t(x)|^2$ for times $t=0,2,\cdots ,14$.
This is shown by solid black lines appropriately offset
vertically. Short-dashed lines on the top show $\ln(\rho_t(x))/4$
(offset, divided by 4 for convenience) for the longest times
$t=10,12,14$. Results compare extremely well with the classical
calculation in Fig.~\ref{fig_5.3}. To illustrate this, we show for
the same classical Gaussian distribution as in Fig.~\ref{fig_5.3}
the distribution $p(x,t)$ defined in Eq.~(\ref{rho_xt_cl_def}):
histogram at the bottom-left shows $p(x,t=0)$, histogram at the top
shows $\ln(p(x,t=14))/4$.
The only ``true quantum'' effects visible in this figure are the small
interference steps at the left wing of the sharpest peak $t=6$.
 }
\end{figure}

The evolution of an initially compact distribution described by
Eq.~(\ref{rho_tilde}) is shown in Fig.~\ref{fig_4}, which reveals
the essence of scattering of distributions by the smooth barrier.
In terms of individual classical trajectories, the parabolic barrier
works as a perfect beamsplitter with all the trajectories with
positive energy ($\eps>0$) being transmitted and all the
trajectories with negative energy ($\eps<0$) being reflected.
However, in terms of the evolution of a probability distribution, there is no splitting
into two compact transmitted and reflected distributions. This is
simply impossible in the case of the stretching and squeezing phase
space dynamics within the beamsplitter region. If the average energy of the distribution is close
to $\langle\eps\rangle =0$, all what one would see after a long
time is an ensemble of particles sitting on the top of the barrier and
uniformly (exponentially in time) expanding in both directions.

Importantly, although the behavior shown in Fig.~\ref{fig_4} is purely classical, it describes
correctly the evolution of the quantum mechanical Wigner
function~\cite{Husimi1953,Moyal}, as we discussed
above. The Wigner distribution carries the information about the quantum tunneling  via the uncertainty principle as it contains a superposition of states both above and below the barrier~\cite{Heim2013}.
Pursuing the line of reasoning similar to the one that led to
Eq.~(\ref{rho_tilde}) we find a more complicated formula relating
the densities in phase space at times $t$ and $0$  in the
original $x$ and $p$ coordinates
 \begin{align}\label{rho_xpt}
&\rho(x_t,p_t,t)= \rho(x_0(x_t,p_t),p_0(x_t,p_t),0)\\
&= \rho(x_tc_\lambda -\fr{p_t
s_\lambda}{m\lambda},p_tc_\lambda-m\lambda x_ts_\lambda ,0) \ ,
\nonumber
 \end{align}
where $c_\lambda =\cosh \lambda t$ and $s_\lambda =\sinh\lambda
t$. The coordinates $x_t$, $p_t$ describe the
particle at a time slice $t$ and $x_0(x_t,p_t)$, $p_0(x_t,p_t)$ is
the initial point of the trajectory found as a function of the
end-point. The expression Eq.~(\ref{rho_xpt}) is still exact.


For direct comparison with the evolution of quantum mechanical
wave packet it is convenient to introduce the classical density in
the coordinate space
 \bq\label{rho_xt_cl_def}
\rho(x,t)=\int \rho(x,p,t)dp \ .
 \ee
Replacing integration over coordinates at current time $t$ by that
at $t=0$ we write
 \begin{align}
\rho(x,t)&=\int\delta(x-x_t)\rho(x_t,p_t,t)dx_tdp_t \\
&=\int\delta(x-x_t(x_0,p_0))\rho(x_0,p_0,0)dx_0dp_0 \ .\nonumber
 \end{align}
From this we derive
 \bq 
\rho(x,t)=
\int \rho\left( \fr{m\lambda \, x-p\sinh\lambda t
}{m\lambda\cosh\lambda t}, p, 0\right) \ \fr{dp}{\cosh \lambda t}
\ .
 \ee
This formula is exact for the classical motion induced by the Hamiltonian Eq.~(\ref{H_barrier}). However, as long as the classical
distribution is further stretched and becomes effectively
one-dimensional, the particle 
inside the barrier is
drastically simplified
 \begin{align}\label{rho_0}
\rho(x,t\gg &1/\lambda)=e^{-\lambda t} f(e^{-\lambda t}x) \ ,\\
f(x)&= 2\int \rho(2 x -{p}/{m\lambda},p,0)dp \ . \nonumber
 \end{align}
Here we introduced a scaling function $f$ such that at large times
$e^{\lambda t}\gg 1$ the evolution of the density reduces to a
simple stretching of $f(x)$.

Formula~(\ref{rho_0}) describes effectively the expansion in time
of the mono-energetic distribution of the ``chain'' of particles. In
Fig.~\ref{fig_4} this corresponds to a line distribution for the
long times along the unstable direction in phase space. The
energy width of the initial distribution $\rho(x,p,0)$ can be
ignored since after long times, $t\gg 1/\lambda$, only the
particles with negligibly small energies, $\eps\approx 0$, are
still present near the top of the barrier. Other particles by the
time $t\gg 1/\lambda$ escape so far away from the $x\approx 0$
region that their energies may be ignored compared to the
magnitude of the potential. Individual trajectories for particles
constituting the distribution Eq.~(\ref{rho_0}) have a simple
exponential form
 \bq\label{x<Lbarrier}
x(t)=x_0 e^{\lambda t} \ .
 \ee
The accelerated exponential divergency of such trajectories,
specific for the Hamiltonian Eq.~(\ref{H_barrier}) with bottomless
reverse-parabolic potential, could not last forever. In a real
system, the potential in Eq.~(\ref{H_barrier}) at some distance
$|x|> L_{\rm barrier}$ reaches a minimum and flattens. The particles
after reaching this region continue to fly with the constant
velocity $v\approx L_{\rm barrier}\lambda$. Trajectory
Eq.~(\ref{x<Lbarrier}) for $x>L_{\rm barrier}$ (or for
$t>\ln(L_{\rm barrier}/x_0)/\lambda$) transforms into
 \bq\label{x>Lbarrier}
x(t)=L_{\rm barrier} +v\, (t-\ln(L_{\rm barrier}/x_0)/\lambda) \ ,
 \ee
where the large logarithm is found up to a constant $\sim 1$
depending on the details how the barrier transforms from the parabolic
top to the flat bottom.

The density distribution for particles (either reflected, or
transmitted) described by Eq.~(\ref{x>Lbarrier}) is related to the
initial phase space density $\rho(x,p,0)$ via the function $f(x)$
defined in Eq.~(\ref{rho_0}) as 
 \bq\label{rho_xt}
\rho(x,t)=(\lambda/v)R\, e^{(x-vt)\lambda/v} f(R\,
e^{(x-vt)\lambda/v}) \ ,
 \ee
where $R=\text{const}\times L_{\rm barrier}$ with some model dependent
constant of order unity.

In Fig.~\ref{fig_5.3}, we show the evolution of the classical
distribution of particles found numerically for the realistic model
barrier $V(x)= A^2/\cosh(x/A)$ supporting all the qualitative
features of our theoretical picture. (Realistic here means a potential with a wide parabolic top, $L_{\rm barrier}\sim A\gg 1$, Eq.~(\ref{x>Lbarrier}), reaching a flat bottom $U=0$ for $|x|\gg A$.)
The figure shows how the compact Gaussian distribution approaching the potential barrier becomes effectively a single elongated line distribution collapsing to the separatrix trajectory $p_s(x)$ with the center pinned to the unstable stationary point. By plotting the displacement density on a logarithmic scale $x,p\rightarrow x,\ln(p-p_s(x))$ 
we show both -- the  flying head of the distribution Eq.~(\ref{rho_xt}) and the universal exponential tail. 

Equation~(\ref{rho_xt}) shows the  shape of the entire particle
distribution after passing the beamsplitter. One may place a
detector at some point $x$ far away from the beamsplitter and
measure the probability for the particle to be found there after a time
$t$. At long times, the function $f$ in Eq.~(\ref{rho_xt})
saturates at $f(0)$. One expects here to measure the
universal exponential tail of the rare events with large delay
time $\sim e^{-\lambda t}$ with the exponent $\lambda$ determined
by the curvature of the potential barrier. We stress that this is the tail of the probability
to find {\it both} electrons still in the beamsplitter. 

\subsection{Quantum mechanical treatment of a bound molecule at the beamsplitter}

For a pure quantum mechanical ensemble, a uniformly decaying solution \eqref{rho_0} of time evolution can also be obtained as a quasi-stationary solution to the time-dependent Schr\"{o}dinger equation
 with the Hamiltonian
Eq.~(\ref{H_barrier}) as
 \bq\label{psi(t)}
i\hbar\dot{\psi}_\lambda=H_{\rm barrier}\psi_\lambda \ , \
\psi_\lambda =e^{im\lambda x^2/\hbar -\lambda t/2} \ ,
 \ee
which is nothing more than the first quasi-stationary state in a parabolic potential with
the imaginary energy $\eps=i\hbar\lambda/2$. The density produced
by this decaying quasi-stationary state
$|\psi_\lambda|^2=e^{-\lambda t}$ coincides with the
classical density Eq.~(\ref{rho_0}) for $x e^{-\lambda t}\ll 1$.

On the other hand, the classical distribution $\rho(x,p,t)$ for
long times, when Eqs.~(\ref{rho_0},\ref{rho_xt}) apply,
becomes effectively a line in phase space. Such a density may be
produced by a quantum mechanical wave function in the semiclassical approximation,
 \bq\label{Hamilton_Jacobi}
\psi(x,t) =\sqrt{\rho(x,t)} \, e^{iS(x,t)/\hbar} \ ,
 \ee
with an action $S$ found from the classical Hamilton-Jacobi
equation. $\psi(x,t)$ in Eq.~(\ref{Hamilton_Jacobi})
coincides with the quasi-stationary solution $\psi_\lambda(x,t)$ in 
Eq.~(\ref{psi(t)}) at distances
$|x|\lesssim\sqrt{\hbar/m\lambda}$, where $S=m\lambda
x^2$ and $\rho =e^{-\lambda t}$, even though the formal semiclassical
expansion is not expected to be valid at such distances.

To illustrate the accuracy of the developed approximations for the wave packet propagation
through the beamsplitter we performed fully quantum mechanical
simulations for the barrier potential $V(x)= A^2/\cosh(x/A)$. The
resulting density $\rho_t(x)$ as well as the logarithm of the density $\ln(\rho_t(x))$ are shown in Fig.~\ref{fig_5.2}.

We note  that what is called in this section coordinate $x$
and momentum $p$ (either classical or quantum) become operators or
expectation values of two (non-commuting) guiding center
coordinates $X_g$ and $Y_g$ for the center of mass of the composite two-electron particle drifting in the strong
magnetic field, described by the Hamiltonian $H_{\rm cm}^{\rm eff}$, Eq.~(\ref{H_cm}).

The main result of this section is the prediction that for a generic
ensemble of electrons approaching the beamsplitter with
energies close to the pinch off, a fraction
$\sim e^{-\lambda t}$ of the electron pairs will stay there for any long time $t$
with the Lyapunov exponent $\lambda$ determined by the curvatures
of the potential barrier. If one would send two non-interacting
electrons to the beamsplitter, the probability to find them both
still inside the device decay much faster, $\sim e^{-2\lambda t}$ since these are uncorrelated events. 
However, if the two electrons initially form a
bound molecular state, which we introduced in the previous section,
the probability for the pair to stay long at the QPC still decays in a
single-particle way~$\sim e^{-\lambda t}$.


\section{HOM: Quantum tunneling into the molecule}\label{section6}


Using the HOM setup offers an advantage compared to
the HBT setup since colliding electrons from two  independent sources enables greater degree of control over the incoming distributions~\cite{Ubbelohde2022,Fletcher2022}. 
In this section, we describe the important theoretical aspects of the electron collision experiment at the beamsplitter, including the role 
of two-electron anti-bound states.

\subsection{Long-time limit  of two electron density  at the beamsplitter
\label{secVA}}

The space of parameters enumerating distinct classical scattering
processes is three-dimensional as there are initial energies of the two
electrons $i=1,2$, energies $\eps_1$ and $\eps_2$, and the difference between the injection
times, $\Delta t_{\text{in}}$. Interaction effects are obviously most relevant
for the simultaneous injection, $\Delta t_{\text{in}}=0$. Consequently, we
start by considering two electron wave packets $\psi_1(x_i,t)$,
$\psi_2(x_i,t)$ with energies $\approx\eps_1, \approx\eps_2$
injected simultaneously into the opposite arms of the beamsplitter, see Fig.~\ref{fig_1}.

We again are interested in the drift motion of the two electrons and consequently use guiding center coordinates $x_i$ and $y_i$. However, only
one of the electrons' non-commuting (see Eq.~(\ref{canonical}))
guiding-centers coordinates $x_i$ or $y_i$ may be treated as a
canonical ``coordinate''. We so far have chosen the ``coordinate'' to be the ``valley of the saddle''
direction $x$, Eq.~(\ref{saddle_point}), which is the direction in which the current flows through the beamsplitter.  Equivalence of the 
coordinates is restored if one considers the density defined in terms of the Wigner function
 \begin{align}
&\rho_i(\vec r_j,t)= \nonumber \\
\fr{1}{\pi l_B^2}
\int\psi_i^*(x_j+x'_j&,t)\psi_i(x_j-x'_j,t)e^{2iy_jx'_j/l_B^2} dx'_j \, 
. \label{Wigner}
 \end{align}
The two electrons injected into the beamsplitter are initially uncorrelated and described by an antisymmetrized product of their individual wave functions
\begin{align}\label{injection}
\psi(x_1,x_2,t)&=  \nonumber \\
\fr{1}{\sqrt{2}}(\psi_1(x_1,t)\psi_2(x_2,t)-\psi_2(x_1,t)\psi_1(x_2,t))
\, .
\end{align}
Here, we assume   fully spin-polarized electrons and suppress the spin indices.

The corresponding density can be computed in analogy to Eq.~\eqref{Wigner} as
\begin{gather} 
   \rho(\vec r_1,\vec r_2,t)=\frac{1}{\pi^2 l_B^4}\iint\psi^{\ast} (x_1^{}+x_1',x_2^{}+x_2',t)   \nonumber \\ \times \psi(x_1^{}-x_1',x_2^{}-x_2',t) e^{2i (y_1^{} x'_1 +y_2^{} x'_2)/l_B^2} dx_1'dx_2'. \label{Wigner2}
\end{gather}

Describing electrons as independent in terms of coordinates $x_1$, $x_2$ in Eq.~(\ref{injection}) works only at sufficiently early times, as long as the interaction may be neglected. This should be
modified when the particles approach each other. As we discussed at length in Section~\ref{section3}, in the region where  the
saddle point potential approximation Eq.~(\ref{saddle_point}) can be
applied, the  effective Hamiltonian for the guiding centres of the two electrons
decouples into a sum of two Hamiltonians describing the motion of
the center of mass, $H_{\rm cm}^{\rm eff}$ in Eq.~(\ref{H_cm}), and
the relative coordinate, $H_{\rm r}^{\rm eff}$ in Eq.~(\ref{H_r}). 
Therefore, there exists a class of two-electron states in the beamsplitter region that remain a product of two independent wave functions at any time if expressed in terms of $x=x_1-x_2$ and $X=(x_1+x_2)/2$,
 \bq\label{psi_sep}
\psi(x_1,x_2,t)=\psi_{\rm r}(x,t)\,
\psi_{\rm cm}(X,t) ,
 \ee
where $x$ and $X$ are the relative and center-of-mass coordinates, respectively.
Fermi statistics of spin-polarized electrons is accounted for by the symmetry condition $\psi_{\rm r}(x,t)=-\psi_{\rm r}(-x,t)$. Note that here $x$ and $X$ refer to guiding center coordinates of the relative and center of mass motion, respectively (i.e. $x\equiv x_g$ and $X\equiv X_g$).

The exact electron density in Eq.~\eqref{Wigner2} in this case also
factorizes ($\vec r=\vec r_1-\vec r_2$, $\vec R=(\vec r_1 +\vec
r_2)/2$) 
 \bq\label{rho_sep}
\rho(\vec r_1, \vec r_2,t)
=\rho_{\rm r}(\vec r,t)\,
\rho_{\rm cm}(\vec R,t)  .
 \ee
Here, the relative coordinate $\rho_{\rm r}$ and center of mass
$\rho_{\rm cm}$ densities are found from the corresponding wave functions $\psi_{\rm r}$ and $\psi_{\rm cm}$ using  Eq.~(\ref{Wigner}) with $l_B^2$ replaced by $2l_B^2$ and $l_B^2/2$, respectively, to account for commutation relations (\ref{commutation}). The exchange symmetry, $\psi_{\rm {r}}(x,t)= \pm\psi_{\rm {r}}(-x,t) $, implies $\rho_{\rm{r}}(\bm{r},t)=\rho_{\rm{r}}(-\bm{r},t)$, independent of the sign  (which is `$-$' in our case).

\begin{figure}
\includegraphics[width=7.2cm]{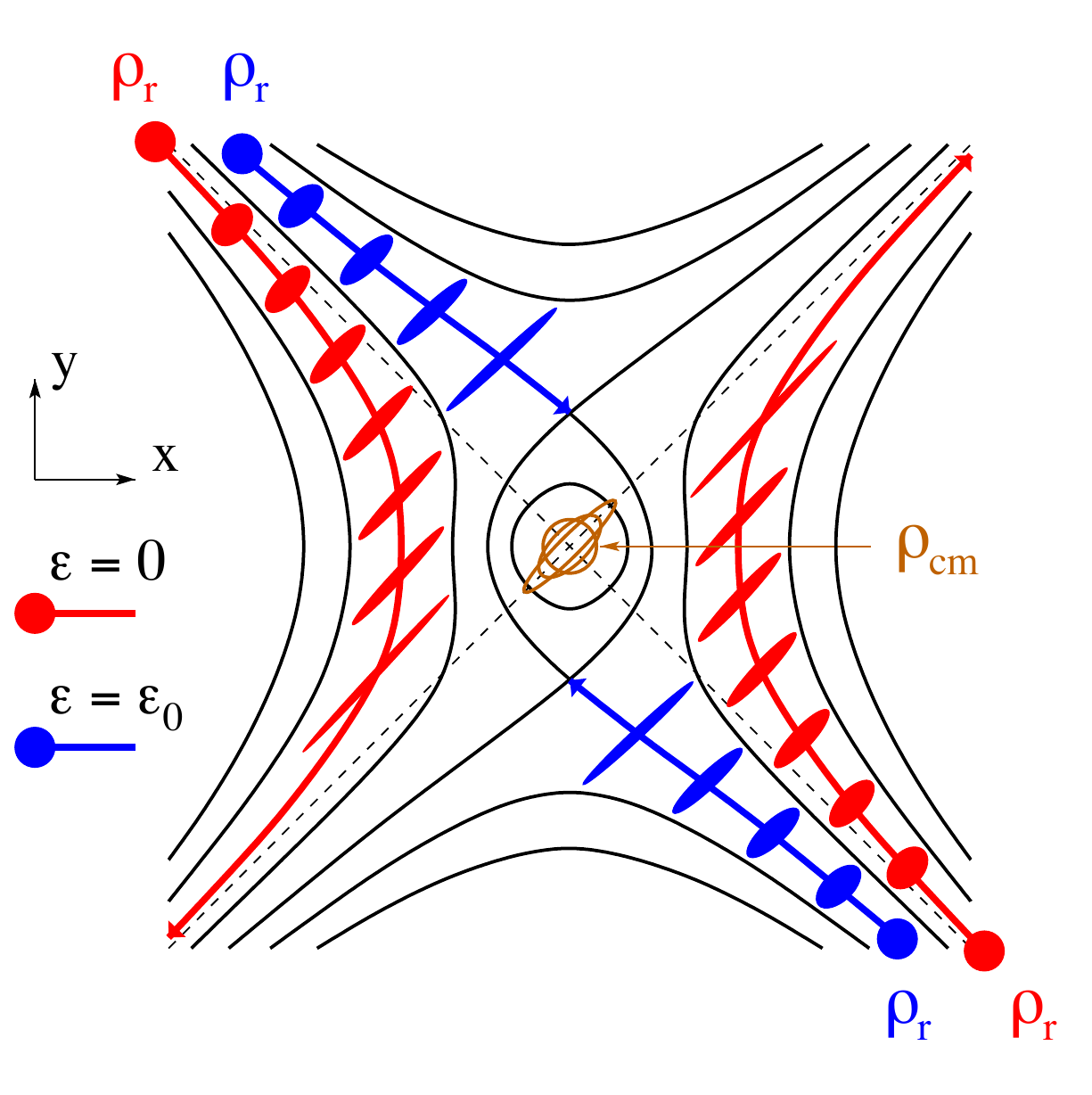} 
\caption{\label{fig_6} Simultaneous injection of two Gaussian
distributions of electrons in the HOM setup with energies $\eps$ close to
$0$~(red) and $\eps_0$~(blue), cf. Eq.~(\ref{eps0}). The
figure shows the central trajectories ($\eps\equiv 0$, $\eps\equiv
\eps_0$) and a schematic evolution of the distribution around them
for the relative coordinate density $\rho_{\rm r}(\vec r,t)$,
Eq.~(\ref{rho_sep}).  The solid black lines show the contour map of the
effective Hamiltonian $H_{\rm r}^{\rm eff}$, Eq.~(\ref{H_r}). The evolution of the
distribution of center of mass trajectories $\rho_{\rm cm}(\vec
R,t)$ is shown schematically in the center (brown). Since
we consider the simultaneous equal-energies injection, the center
of $\rho_{\rm cm}(\vec
R,t)$ stays at the stationary point $\vec
R(t)\equiv 0$.}
\end{figure}

Solutions of the time dependent 
Schr\"{o}dinger equation for the guiding centers of the form Eq.~(\ref{injection}) and
Eq.~(\ref{psi_sep}) may, but don't have to coincide even for the large
separation between electrons, where their interaction may yet be
neglected. 
Nevertheless, there exists an interesting family of
solutions for which the factorization Eq.~\eqref{psi_sep} holds for the initial state Eq.~\eqref{injection} and hence is preserved during subsequent time evolution.
Consider $\psi_i(x_i,t)$ as a Gaussian in $x_i$ with some time-dependent, possibly complex, coefficients. A sum of two at most quadratic single-variable polynomials in $x_1$ and $x_2$ is equal to the sum of single-variable polynomials in $x$ and $X$ if and only if the coefficients in front $x_1^2$ and $x_2^2$ are equal. 
When this condition is true for $\ln \psi_i$, the initial state Eq.~\eqref{injection} factorizes with $\psi_{\rm r}(x,t)=[\psi_1(x/2,t)\psi_2(-x/2,t)-\psi_1(-x/2,t)\psi_2(x/2,t)]/\sqrt{2}$ and $\psi_{\rm cm}(X,t)=\psi_1(X,t)\psi_2(X,t)$. The initial wave function of the relative coordinate in this case is a cat state (superposition of well-separated Gaussians).
If these incoming  Gaussian wavepackets are also tuned to be symmetric with respect to the origin,
$\psi_1(x,t)=\psi_2(-x,t)$, then $\psi_{\rm r}(x,t)=[\psi_1^2(x/2,t)-\psi_1^2(-x/2,t)]/\sqrt{2}$ and $\psi_{\rm cm}(X,t)$ is a Gaussian centered at the origin $X=0$.
The corresponding initial density $\rho_{\rm r}$ is a sum of two Gaussians (sans fine oscillations in $\rho_{\rm r}$ on the scale much smaller than $l_B$, which are characteristic to cat states and encode the sign of the exchange statistics) and $\rho_{\rm cm}$ is a Gaussian too. 
We
will analyze in details the evolution of such a special distribution
in Fig.~\ref{fig_6}.

In the most general case, the two-particle wave function in the beamsplitter region can be expressed via the multi-component generalization
of Eq.~(\ref{psi_sep}) 
\begin{align}\label{psi_entangled}
\psi(x_1,x_2,t)
=\sum_i &c^i\psi_{\rm r}^i(x,t)\,
\psi_{\rm cm}^i(X,t)  \, ,
 \end{align}
which is possible with some appropriate choice of
$\psi_{\rm r}^i, \psi_{\rm cm}^i$ as Eq.~\eqref{psi_entangled} is a Schmidt decomposition of the two-electron state after the coordinate transformation.
The separation of variables in the quadratic Hamiltonian ensures that the coefficients $c^{i}$ are constants of motion.

The general wave function Eq.~\eqref{psi_entangled} simplifies drastically in the special interesting
regime of long times and the center of mass coordinate $X$ staying close to the top of the barrier. In this regime,
individual center of mass wave functions all behave like the quasi-stationary solution
Eq.~(\ref{psi(t)}) discussed at the end of the previous section
$\psi_{\rm cm}^i(X,t)\rightarrow \delta^i \exp[im\lambda X^2/\hbar -\lambda t/2]$ with some coefficient $\delta^i$.
Thus effectively, here, Eq.~\eqref{psi_entangled} became a single product wave function Eq.~(\ref{psi_sep}) with
 \begin{align} 
&\psi_{\rm cm}(X,t)=e^{im\lambda X^2/\hbar -\lambda t/2} \ , \nonumber \\
&\psi_{\rm r}(x,t)=\sum_i c^{i} \delta^i \psi^{i}_{\rm r}(x,t) \ . \label{CM_uniform}
 \end{align}
Here, for the saddle point potential Eq.~(\ref{saddle_point}) and
magnetic field $B$ we have $m\lambda=2(eB)\sqrt{A/C}$ and
$\lambda=(1/eB)\sqrt{AC}$. 
This $\psi_{\rm cm}(X,t)$ describes the longest surviving quasi-stationary state of the inverted parabolic potential,  Eq.~\eqref{psi(t)}. Further evolution of the relative coordinate wave function $\psi_{\rm r}(x,t)$ 
is governed by the effective Hamiltonian $H^{\rm eff}_{\rm r}$ Eq.~(\ref{H_r}), which we will consider later in Sec.~\ref{sec:ModelHamiltonian}.

\subsection{Classical scattering at the beamsplitter\label{secVB}
}

Having discussed the effective universality of the decomposition into the relative and center of mass coordinates,
Eqs.~(\ref{psi_sep}, \ref{rho_sep}), we may now proceed with
considering the interaction effects in the two-electron HOM
experiment. In Fig.~\ref{fig_6}, we show (schematically) two
small similar 
distributions $\rho_1(\vec r_1,t)$
and $\rho_2(\vec r_2,t)$ with the same mean energy $\eps_1=\eps_2=\eps$ injected
simultaneously through the opposite corners of the beamsplitter.
In fact we show two pairs of such distributions --- blue and red
--- for the reasons explained below. We assume that it is possible
to create the distributions with the spatial dimensions
small compared to the size of the beam splitter
($L_{\rm barrier}$ of the previous section). 

For our first example,  (red in Fig.~\ref{fig_6}) we placed the
centers of both incoming electrons distributions $\rho_1$ and
$\rho_2$ at the separatrix, which is the straight equipotential
 line
 of $V(x,y)$ separating the non-interacting
single-electron trajectories with topologically different
outcomes. For this choice the fate of the central trajectory of the
wave packet/distribution is classically undetermined and effects
of quantum mechanics 
became most pronounced.
However, long-range interaction
renders this former critical point for the relative coordinate of the two electrons trivial. Electrons' repulsion
in the relative coordinate Hamiltonian increases the
effective height of the barrier, from which the electrons are now
simply reflected without spending much time inside the
beamsplitter and thus not changing qualitatively their
distribution. 
The process may be viewed as 
 a classical particles anti-bunching with none of the 
 electrons
 crossing the
beamsplitter.

The new interaction-induced critical point is revealed  if one
raises the energy of incoming electrons such that they would be
able to enter the beamsplitter and approach (asymptotically) the
saddle point at $y=\pm y_0$ of the joint potential $V(x,y)/2+e^2/(\kappa r)$.  
Corresponding trajectory for the center of the
electrons distribution is shown blue in~Fig.~\ref{fig_6}.

The straightforward way to reach the 
critical point in the presence of the Coulomb interaction 
is to send two electrons 
simultaneously to the beamsplitter each with an extra energy 
 \bq\label{eps0}
\eps_0 =\fr{3 (Ce^4/\kappa^2)^{1/3}}{4\cdot 2^{1/3}} \ .
 \ee 
Here, $C$ is the transverse curvature of the saddle point potential,
Eq.~(\ref{saddle_point}), $e$ is the electric charge and $\kappa=4 \pi \epsilon \epsilon_0$ is the dielectric constant. 
The critical
energy $\eps_0$, Eq.~(\ref{eps0}), is found as a half of the
saddle-point energy of the effective Hamiltonian $H_{\rm r}^{\rm
eff}-\hbar\omega_L/2$, Eq.~(\ref{H_r}), describing the relative coordinate $\vec r
=\vec r_1 -\vec r_2$ dynamics. 
The value of $\varepsilon_0=3 U/4$ where $U$ parametrizes the Coulomb potential~\cite{Pavlovska2022}, $e^2/(\kappa r)=U y_0/r$,  in the experiment of Ref.~\onlinecite{Ubbelohde2022} has  been estimated to be about $\varepsilon_0 \approx \SI{2}{m\electronvolt}$.


If two electrons are sent to the beamsplitter with the time shift
$\Delta t_{\text{in}}$, they still may stop at the same stationary critical
point. However, the excess energy $2\eps_0$ necessary to reach it
must now be split unevenly between the electrons. Interestingly, in case of
the sufficiently large saddle-point region, Eq.~(\ref{saddle_point}), the two energies
$\eps_1(\Delta t_{\rm in})$ and $\eps_2(\Delta t_{\rm in})$ may be found analytically in a simple form
 \bq\label{eps12}
\eps_1(\Delta t_{\rm in})=\fr{2\eps_0}{1+e^{\lambda\Delta t_{\rm in}}} \ , \
\eps_2(\Delta t_{\rm in})=\fr{2\eps_0}{1+e^{-\lambda\Delta t_{\rm in}}} \ .
 \ee
Here, the pair's center of mass materially moves along its critical trajectory
approaching the stationary point only asymptotically, 
which is not described by
Fig.~\ref{fig_6}. 
Knowledge of $\eps_1(\Delta t_{\rm in})$, $\eps_2(\Delta t_{\rm in})$ allows us to find $\eps_0$, Eq.~(\ref{eps0}), and 
$\lambda =(1/eB)\sqrt{AC}$ and thus infer the curvatures of the saddle point $A,C$, Eq.~(\ref{saddle_point}).

For $\lambda\Delta t_{\rm in} \gg
1$ the first electron in Eq.~(\ref{eps12}) is sent to the beamsplitter with the
energy just marginally above the half-transmission and
waits at the saddle point at the origin for the energetic
partner from the other side. Then they relax at two
emerging saddle points.

Derivation of Eq.~\eqref{eps12} requires only the knowledge of electrons trajectories when they are
well separated and don't interact (while both are already present inside the parabolic beamsplitter).
The electrons' interaction enters this result only through the total electrons' energy $2\eps_0$.
The simple derivation of Eq.~\eqref{eps12} is given in the Appendix.
More generally, the classical scattering problem inside the saddle-point beamsplitter potential Eq.~(\ref{saddle_point})
can be reduced to quadratures~\cite{Pavlovska2022,Ubbelohde2022} for any combination of the incoming electron energies $\varepsilon_{1\text{in}}$ and
$\varepsilon_{2\text{in}}$ and the incoming time shift $\Delta t_{\text{in}}$. Equation \eqref{eps12} can be seen as a special case of the analytic scaling relations \cite{Pavlovska2022} that map half-transmission thresholds from  coincident arrival ($\Delta t_{\text{in}}=0$) to a finite $\Delta t_{\text{in}}$.

As already discussed in the context of Fig.~\ref{fig_6},
the injection of two electrons at 
the non-interacting
{half-transmission energy} $\varepsilon_{1\text{in}}=\varepsilon_{2\text{in}}=\eps =0$ is non-critical, 
{\it i.e.} there is no change of the trajectory topology upon a small variation of initial conditions here.  
Nevertheless, the presence of the interaction leads to the interesting 
and potentially measurable effects for this choice of initial conditions (which can be tuned without synchronizing the sources).
In Fig.~\ref{fig_6}, by red lines we show the trajectories for simultaneous ($\Delta t_{\text{in}}=0$) injection of two $\eps=0$ electrons. Consider now, what happens if the two 
electrons are injected still both at $\eps=0$ but with a finite relative time delay $\Delta t_{\text{in}}$.
In this setup we expect two clear signatures of interaction. 
First, for a pair of electrons each injected at non-interacting half-transmission but with a relative time delay, the center of mass does not stay at the stationary position $X,Y=0$ of the Hamiltonian $H_{\rm cm}^{\rm eff}$ Eq.~(\ref{H_cm}), but approaches it asymptotically at large times, $t\rightarrow \infty$.
Thus the time delay between two electrons distributions measured after the reflection will be reduced, $\Delta t_{\text{out}} <\Delta t_{\text{in}}$, meaning electrons are effectively synchronized by the interaction (time domain bunching).
Second, while the interaction tends to time-synchronize the electrons, they exchange the energy in an elastic collision, 
with the electron reaching the beamsplitter first (second) gaining (loosing) energy.

Using the general solution for energy transfer [Eq.~(26) of Supplementary note to Ref.~\onlinecite{Ubbelohde2022}] 
we can~\footnote{In terms of the dimensionless function used in Refs.~\onlinecite{Pavlovska2022,Ubbelohde2022}, one has $\Phi_k= E_{+}/(2 \gamma_0 \varepsilon_0)$ for $|\Phi_k| \ll 1$ with $E_{+}=\eps [1+\cosh(\lambda \Delta t_{\text{in}})]$ if $\eps=\varepsilon_{1\text{in}}=\varepsilon_{2\text{in}}$.} 
compactly express the corresponding outgoing energies, 
\begin{align} \label{eq:energytransfer}
    \varepsilon_{1\text{out}}=-\varepsilon_{2\text{out}}=\gamma_0 \, \varepsilon_0 \tanh(\lambda \Delta t_{\text{in}}/2) \, ,
\end{align}
where $\gamma_0= 2 [2 A/(A+C)]^{1/3}/3$ is a dimensionless constant of order 1.  
The functional dependence of $\varepsilon_{1\text{out}},\varepsilon_{2\text{out}}$ in Eq.~\eqref{eq:energytransfer} on $\Delta t_{in}$ is derived in the Appendix, while finding the overall coefficient
in Eq.~\eqref{eq:energytransfer} requires the methods used in~\cite{Pavlovska2022}. 
We see that $\varepsilon_0$ given by Eq.~\eqref{eps0} sets the typical scale of non-critical energy gain/loss for time separations $\Delta t_{\text{in}}$ larger than the scale set by the inverse Lyapunov exponent $\lambda^{-1}$.

The corresponding synchronization effect can be quantified 
in terms
of the time shift between the incoming and outgoing electrons, $\Delta t_{\rm in}$ and $\Delta t_{\rm out}$, as defined by the asymptotics of the trajectories incoming from opposite sources and outgoing to the opposite detectors,
 \begin{align}
x_1(t)/x_2(t)& = \begin{cases}
-e^{-\lambda\Delta t_{\rm in}} \, ,  & t\rightarrow -\infty \\  
-e^{+\lambda\Delta t_{\rm out}} \, ,  & t\rightarrow +\infty 
\end{cases} \, . 
 \end{align}
Using Eq.~(20) of Ref.~\onlinecite{Pavlovska2022} for the asymptotics of outgoing trajectories, we obtain a compact relation
\begin{align} \label{eq:timedelay}
\tanh (\lambda \Delta t_{\rm out}/2) & = -\frac{\eps}{2 \gamma_0 \varepsilon_0} \sinh (\lambda \Delta t_{\rm in}/2) \, .
\end{align} 
The r.h.s.\ of Eq.~\eqref{eq:timedelay} vanishes for $\eps=0$ and remains a good approximation up to $|\varepsilon|  \cosh (\lambda \Delta t_{\text{in}}/2) \lesssim \varepsilon_0 \gamma_0$.  This implies that $ |\Delta t_{\text{in}}|$ up to $\sim \lambda^{-1} \ln  (\gamma_0 \varepsilon_0/|\eps|)$ can be compensated by interaction to achieve synchronization within  the characteristic time $\lambda^{-1}$ of the beamsplitter, 
$\Delta t_{\rm out} \lesssim \lambda^{-1}$ 
\footnote{Note that a change in energy leads to a logarithmic time shift even for one non-interacting electron as this is a basic dispersion property of an energy-selective beamsplitter (see Section~\ref{section5} above and Sec.~III-C of Ref.~\onlinecite{Pavlovska2022}).}.

In the examples above we have focused on the trajectories of the centers of sufficiently narrow statistical distributions $\rho_{1(2)}$. In the latest experiments~\cite{Ubbelohde2022,Fletcher2022} the uncertainty of the incoming energy and time distributions, as characterized by energy-time tomography~\cite{Fletcher2019} of the sources, was found to be comparable to the  characteristic scales $\mathcal{\varepsilon}_0$ 
and $\lambda^{-1}$ of the interactions on the beamsplitter. Therefore, a statistical approach~\cite{Ubbelohde2022}
averaging over classical outcomes~\cite{Pavlovska2022} had to be developed for quantitative analysis of collision statistics.
Both experiments \cite{Ubbelohde2022,Fletcher2022} have demonstrated capabilities relevant to the specific examples discussed in this subsection such as the ability to 
tune the relative delay  $\langle \Delta t_{\text{in}} \rangle$ between the centers of the time distributions of the two electrons.
In the experiment of Ref.~\onlinecite{Fletcher2022}, both electrons were kept at the same energy, 
while the height of the beamsplitter barrier was varied. 
In Ref.~\onlinecite{Ubbelohde2022}, the energy of one electron was kept at the non-interacting half-transmission, while the  energy of the second injected electron was scanned.

A distinctive feature of the experiment by Ubbelohde et al.~\cite{Ubbelohde2022} is the access to full counting statistics of on-demand collision outcomes,  including rare events when less than two electrons arrive at both detectors. As the  energy relaxation during propagation along the edge of sample is sensitive to electron energy, the loss signal serves as a proxy for the energy gain or loss  after an elastic collision at the beamsplitter, i.e. $\varepsilon_{1\text{out}}-   \varepsilon_{1\text{in}}=-(\varepsilon_{2\text{out}}-\varepsilon_{2\text{in}}) $. This loss signal has been measured and successfully  modeled  as function of $\Delta t_{\text{in}}$ and $\varepsilon_{1\text{in}}$ in Ref.~\onlinecite{Ubbelohde2022} (see Fig.~4 in the main text and Supplementary note VI). The reported agreement  confirms the energy exchange effect.

\subsection{Quantum dynamics of two electron scattering near critical points of the effective potential\label{sec:ModelHamiltonian}}

In the previous section, we have analyzed the classical dynamics induced by the effective Hamiltonian 
$H^{\rm eff}_{\rm r}(x,y)$ in terms of individual trajectories in phase space of the relative position components $x$ and $y$ as the latter form a pair of canonically conjugate variables. 
Rigorous quantum mechanical treatment, however, is challenging as these non-commuting variables are mixed in the interaction part of the Hamiltonian, $e^2/(\kappa \sqrt{x^2+y^2})$, in a very non-linear manner. Nevertheless, as we show below, good qualitative understanding of quantum mechanical effects in two wave packet collision can be achieved by mapping the problem 
onto a model Hamiltonian which on the one hand reproduces the essential properties of the original effective Hamiltonian, 
but in addition allows for a straightforward numerical solution.

 \subsubsection{Model Hamiltonian for quantum dynamics near the interaction-induced critical points}

\begin{figure}
\includegraphics[width=9.cm]{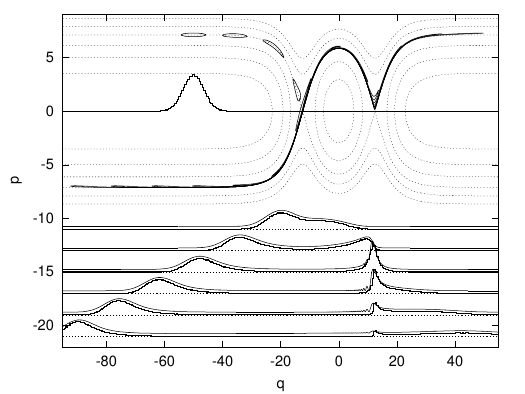} 
\caption{\label{fig_9} 
Phase space dynamics induced by a double-barrier model, Eq.~\eqref{H_Model}, describing the relative motion of two initially well-separated electrons. 
Top: Liouvillian evolution of an initially Gaussian density $\rho(q,p,t)$, traced by snapshots of the half-height contour at $t=0, 2, \ldots, 20$. Constant energy lines (dotted) guide the flow, the distribution is centered on the separatrix (resulting in equal transmission and reflection probabilities). Parameters of the potential \eqref{H_Model} and the initial state \eqref{eq:modelGaussian} are $a=5$, $\delta=2.5$ and  $q_0=-10a$, $p_0=\sqrt{2}a$, respectively. Bottom:
probability  density for $q$ at six selected times $t=10, 12, \ldots 20$ (from top to bottom), computed by projecting classically evolved  phase-space density,
$\rho(q,t)=\int\rho(p,q,t) dp$ (thick lines), 
and by solving the time-dependent Schr\"{o}dinger equation, $| \psi(q,t)|^2$ (thin lines, slightly shifted upwards for comparison).}
\end{figure}

Combination of the Coulomb interaction with the beamsplitter saddle point potential 
in $H^{\rm eff}_{\rm r}(x,y)$ creates two degenerate saddle points leading 
to the division of the $x,y$ plane into four regions of topologically distinct infinite motion and a single region 
of the bound motion hosting quasi-stationary quantum states, see the right panel of Fig.~\ref{fig_3}.
These features are present in the model double barrier 
Hamiltonian, which we choose to be, $y\rightarrow q$, $x\rightarrow p=-i\partial/\partial q$,
 \begin{align}\label{H_Model}
&H^{\rm mod}(p,q)=\frac{p^2}{2} + U(q) \ ,\\
U(q)&=\fr{a^2}{\cosh(q/a-\delta)} +\fr{ a^2}{\cosh(q/a+\delta)} \ .\nonumber
 \end{align}
Our mapping also reverses the sign of the potential, $H^{\rm eff}_{\rm r}(x,y) \rightarrow - H^{\rm mod}(p,q)$, such that the bound motion segment of phase space becomes a bottom, not a top of the energy landscape.
Hence, in this picture, the anti-bound molecular states correspond to quasibound states between the two barriers.
We consider  the barrier width $a$ and the half-distance $\delta$ between the barriers large compared to unity. In this parameter range 
the dimensionless
potential $U(q)$, Eq.~(\ref{H_Model}), has two maxima
at $q=\pm q_m\approx  a \delta$ 
with a quadratic approximation $U= U_0- \lambda^2_I (q \pm q_m)^2/2$ with $U_0 \to a^2$ and $\lambda_I \to 1$ for large $q_m$. 
Hence the classical trajectories  near the energy $U_0$, being the level lines of $H^{\rm mod}(p,q)$, approximate the topology of trajectories of $H^{\rm eff}_{\rm r}(x,y)$,  compare Fig.~\ref{fig_9} (dotted lines in the top panel) and Fig.~\ref{fig_3} (panel on the right).
In this regime  the region of finite motion 
supports a sufficient number of quasi-bound states as its area  $\sim a \delta \gg 1$ and $[p,q]=-i$.

To characterize the original model Eqs.~(\ref{H_cm},\ref{H_r}) in Section \ref{section3} we have introduced two Lyapunov exponents: $\lambda=(1/eB)\sqrt{AC}$,
Eq.~(\ref{alpha_beta}), governing the phase space dynamics for either the single electron or for the center of mass of the electron pair and
$\lambda_{\rm I}=(1/eB)\sqrt{3(A+C)C}$, Eq.~(\ref{gamma_delta}), responsible for the motion
on the relative coordinate plane near one of the interaction-induced saddle points.
In the units of the dimensionless  model Hamiltonian, Eq.~(\ref{H_Model}), we have the corresponding $\lambda_I = -\partial^2 U/\partial q^2|_{q=q_m} \to 1$ for large interbarrier separation $a \delta$. Hence in the numerical examples that follow time is measured in units of approximately $\lambda_I^{-1}$.

\subsubsection{Example I: Squeezing and stretching dynamics near interaction-induced saddle points}

We consider the time evolution of a quantum state initialized at $t=0$ by 
a Gaussian wave function,
\begin{align} \label{eq:modelGaussian}
\psi(q,0)=a^{-1/2} \pi^{-1/4} \exp[-(q-q_0)^2/2a^2 +iqp_0] \, .
\end{align}
The corresponding phase space density, c.f.~Eq.~\eqref{Wigner},  is
$\rho(p,q,0) =\pi^{-1}
\exp(-(q-q_0)^2/a^2-(p-p_0)^2a^2)$.
The initial coordinate $q_0=-10 a $ is taken sufficiently far from the double well ($|q_0| \ll q_m$) and the initial momentum $p_0$ corresponds to incoming energy matching the maximum of a single barrier, $p_0^2/2 \approx U_0$.

First we illustrate the  Liouvillian  phase space dynamics generated by $H^{\text{mod}}(p,q)$, neglecting the anti-symmetrization requirement for clarity. 
Figure~\ref{fig_9} shows the evolution of an ensemble of classical trajectories with initial distribution corresponding to the Wigner representation of $\psi(q,0)$. 
Like previously in Figs.~\ref{fig_5.3} and \ref{fig_5.2} we consider both the density in the phase plane
$\rho(p,q,t)$ and the projected density along one coordinate, $\rho(q,t)=\int\rho(p,q,t) dp$. 
 As the Liouvillian flow  preserves
densities and only changes the shapes of the areas, the resulting stretching and squeezing dynamics can be traced by following the half-height level line of the initial density.

We see that after hitting the first potential barrier the density is
split into the reflected part with energy below $U_0$ and the transmitted part with energy above $U_0$ without forming singular features, similar to the single saddle point in the non-interacting problem, c.f.~Fig.~\ref{fig_5.3}. At later times the probability for a system to stay with the relative coordinate close to the first barrier decays as $e^{-\lambda_{I}t}$.

In contrast, in the vicinity of the second barrier (which corresponds to the 
second interaction-induced saddle point of the relative coordinate Hamiltonian),  a singular long-lived density feature 
accumulates, and then slowly decays. This happens because of the trajectories leaving the first barrier and trying to reach the second one
at larger times became closer and closer to the separatrix and thus are delayed more at the second barrier. 
A simple analysis of stretching-squeezing behavior at two consecutive saddle points shows that 
the probability to find the relative coordinate close to the second stationary point decays with half the exponent  $\sim e^{-\lambda_{I}t/2}$.

Additionally, we compare the statistics of the classical trajectories to the exact quantum mechanical solution for the one-dimensional density $|\psi(q,t)|^2$. The comparison in Fig.~\ref{fig_9} show very good overall agreement, except for oscillatory features that develop between the barriers due to interference (most visible in the bottom panel at $q$ just below $q_m \approx 12.5$).
We scrutinize these uniquely quantum features in our next numerical example below.

 \subsubsection{Example II: Effect of exchange symmetry and formation of a quasi-bound state}

\begin{figure}
\includegraphics[width=9.cm]{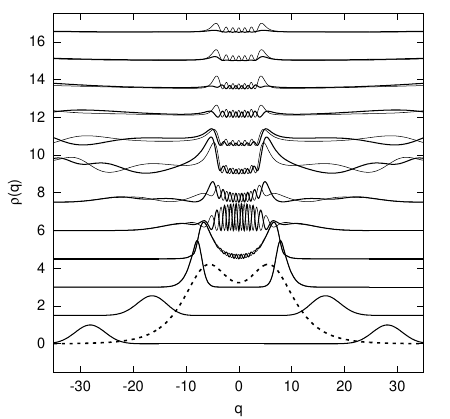} 
\caption{\label{fig_7} 
Evolution of the probability density under the  model Hamiltonian (\ref{H_Model}). The dashed line shows the potential $U(q)$ with parameters $a=4$ and $\delta =1.5$.
The set of twelve continuous-line graphs shows the density profiles for $t=2,4, \ldots 24$ (from bottom to top) of the symmetric ($\rho(q)=|\psi_{+}(q,t)|^2$, thin) 
and antisymmetric
($\rho(q)=|\psi_{-}(q,t)|^2$, thick)
superpositions of initially Gaussian states Eq.~\eqref{eq:Gausssym}, with parameters 
  $q_0=-10a$ and
$p_0=1.051\times\sqrt{2}a$ (central energy is raised slightly to
match the top of the two close barriers). The
top five curves ($t \ge 16$) are multiplied by an extra factor of three.
 }
\end{figure}

In Fig.~\ref{fig_7}, we take a closer look at the snapshots of the probability density $|\psi_{\pm}(q,t)|^2$ as two electrons approach each other and undergo quantum scattering governed by the double-well model Hamiltonian Eq.~\eqref{H_Model}.
This time 
we take into account the symmetry requirements dictated by exchange statistics and consider evolution of both orbitally symmetric ($+$, spin singlet) and anti-symmetric ($-$, spin triplet) states. The properly symmetrized initial state corresponding to the Gaussian Eq.~\eqref{eq:modelGaussian} is
\begin{align}
\label{eq:Gausssym}
\psi_\pm(q,0) \propto &
\bigl\{ \exp [-(q-q_0)^2/2a^2 +iqp_0 ] \\ & \pm
\exp[-(q+q_0)^2/2a^2 -iqp_0] \bigr \} \, .  \nonumber
\end{align}

The sequence of density profiles in Fig.~\ref{fig_7} illustrates the consecutive stages of the time evolution.  We see two initially well-separated wave packets approaching the central
interaction region, where they start to overlap and interfere. This interference 
appears in the form of fast oscillations with smooth envelopes over the maxima and the minima, as the two overlapping components arrive into the central region (near $|q| < |q_m|$)  with opposite and large momenta $p \approx \pm p_0$.  

As the propagating wave reaches the second barrier, a new standing wave pattern emerges, with the envelope of the oscillation minima dropping to almost zero (see $t \geq 10$ traces in Fig.~\ref{fig_7}). This means that both real and imaginary parts of the wave function in the central region nearly vanish and $\psi_{\pm}(q,t)$ at longer times can be represented as an almost real function multiplied by a $q$-independent phase factor. This wave function corresponds to a quasi-bound state with a particular number of nodes between the two barriers. Resonant excitation of such quasi-bound states, illustrated in Fig.~\ref{fig_7}, corresponds to the creation of long-lived electron molecules with well-defined energies and decay width. Quantum tunneling is essential both for excitation and, reciprocally, for the decay of these quasi-bound states with energies below $U_0$.  As the tunneling rates   are exponentially sensitive to the confining potential barrier height, the widths of these resonances will differ by orders of magnitude, hence initially a single wave function with resonant energy close to $U_0$ dominates $\psi_{\pm}(q,t)$. The slowly decaying quasi-stationary density profile corresponding to one of such resonances can be clearly seen for $\psi_{-}(q,t)$ at $t= 20\ldots 24$.   
Crucially, the exchange symmetry of the wave function, $\psi_{\pm}(q,t) = \pm \psi_{\pm}(-q,t)$, plays a crucial role in selecting the excitable resonances, as one can see from the difference between $\psi_{-}$ and $\psi_{+}$ at late times in Fig.~\ref{fig_7}.
This is because for the symmetric double-well potential, $U(q)=U(-q)$,  the parity of the (quasi)-bound one-dimensional wave functions coincides with the parity of the level number, hence the resonance conditions for $\psi_{+}$ and $\psi_{-}$ with the same parameters of the initial state \eqref{eq:Gausssym} are vastly  different.

\subsubsection{Example III: Quantum decay of the molecular states}

\begin{figure}
\includegraphics[width=9.cm]{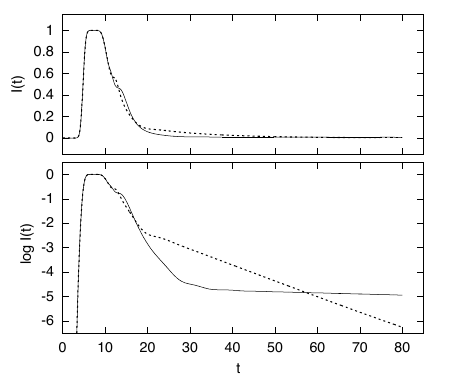} 
\caption{\label{fig_8} 
Probability $I(t)$ of $|q| <q_c= 3.125a$ for the numerical solution illustrated in Fig.~\ref{fig_7}. Here the dashed lines marks the symmetric ($I_{+}$) and the solid lines ---  the antisymmetric ($I_{-}$) case, respectively; bottom panel depicts the same data as the top but on a logarithmic scale.
}
\end{figure}

We further illustrate the excitation and decay of quasi-bound two-electron states in HOM-type two electron collisions by computing the probability $I(t)$ of the pair to remain close as
\begin{align}
   I_{\pm}(t) = \int_{-q_c}^{+q_c} |\psi_{\pm}(q,t)|^2 d q \,.
\end{align}
 For the numerical example analyzed in
Fig.~\ref{fig_7} where $q_m = 5.55$,
 we choose a cut-off distance $q_c = 3.125 a=12.5$ and plot  $I_{\pm}(t)$ in Fig.~\ref{fig_8}. As the selected region is larger than the distance between the barriers and the wave packet width, the initial maximum of $I(t)$ is a plateau close to $1$ as long as the whole wave packet fits into  $|q| <q_0$. The initial decay after the plateau ($t \approx 10 
 \ldots 12$) is rapid as the population is reflected classically at the first ($E<U_0$) or transmitted over the first and then the second  barrier ($E>U_0$), escapes with decay rates $\lambda_{\text{I}}$ and $\lambda_{\text{I}}/2$,  respectively. This corresponds to the relative motion trajectories that separate without  experiencing interference,  as discussed in Fig.~\ref{fig_9} and the accompanying subsection above. In Figure \ref{fig_8}, this ``classical'' stage of evolution is observed up to $t\approx 12$ as two traces of opposite exchange symmetry,  $I_{+}(t)$ and $I_{-}(t)$, remain close to each other.

After the classically reflected and transmitted trajectories have left, the decay of the proximity probability $I(t)$ slows down  and transitions into the asymptotic regime where one expects a piece linear $\ln I (t)$, corresponding to the hierarchy of resonances,
 \begin{align} \label{eq:Iasymp}
   I(t) \propto \sum_n \alpha_n e^{-\gamma_n t} \, .
 \end{align}
As the decay rates $\gamma_n$ for even and odd states are expected to differ substantially, the two traces deviate. A clear exponential behavior at long times ($t>20$ for `$+$' and $t> 50$ for `$-$') is seen in Figure \ref{fig_8}, with rates $\gamma_n=0.064$ and $0.0046$ for the symmetric and anti-symmetric cases, respectively.  These lifetimes are significantly longer than $\lambda_I =0.928$.

Finally, we discuss the long-time behavior of the probability $P(t)$ to find both electrons in the beamsplitter region. This condition requires both (a) the relative distance 
and (b) the center of mass to stay close to the origin.
The probability for (a) is proportional to $I(t)$ in the asymptotic regime of Eq.~\eqref{eq:Iasymp}. The probability for (b) is $\sim e^{-\lambda t}$ as dictated by Eq.~\eqref{rho_0} in Section~\ref{sec:HBTphasespace}.
As the two dynamics are decoupled,
this implies the product of probabilites, i.e.
\begin{align}
   P(t) \sim  e^{-\lambda t} I(t) \sim \sum_n \alpha_n
e^{-\lambda t-\gamma_n t} \, .
\end{align}
This asymptotics agrees with the factorized long-time limit, Eq.~\eqref{CM_uniform}, of a generic initial wave function Eq.~\eqref{psi_entangled}:  the asymptotics of the center of mass wave function, $|\psi_{\text{c.m.}}(x,t)|^2 \sim e^{-\lambda t}$ is driven by the classical Lyapunov exponent $\lambda$, and 
the relative coordinate wave function converges to a superposition of long-lived ($\gamma_n \ll \lambda$) uniformly decaying resonance states, $\psi_{\text{r}}(x,t) \to \sum_n a_n \, \psi_n(x)  e^{-\gamma_n t/2} $.

Hence we conclude that a smooth quadratic beamsplitter acts as a filter, delaying anti-bound states of electron pairs in its vicinity with a characteristic decay of the cumulative probability $P(t) \approx e^{-\lambda t}$ regardless of whether the pairs are formed by tunneling in a collision (HOM geometry, as discussed in this section) or arrive from a potential other source (HBT geometry, discussed in Section~\ref{section5}).
The advantage of the HOM setup could be that there is a clear scenario for controlled generation of the pairs that can serve as an entanglement resource, as we outline below. 

\subsection{Creation of entangled spin-pairs in the time domain\label{secVD}}
The observation of the formation of anti-bound states with different life-times for symmetric (spin-singlet) and antisymmetric (spin-triplet) orbital wave functions provides a way to separate the formation of spin-singlets and spin triplets temporally using the HOM-geometry. We note that the spin-polarized state can only have the antisymmetric orbital wave function (cf. Eq.~(\ref{injection})). It has the total wave function $\Psi(x_1\sigma_1,x_2\sigma_2,t)=\psi(x_1,x_2,t)\chi_{\uparrow}(\sigma_1)\chi_{\uparrow}(\sigma_2)$. Antisymmetry for fermions requires that $\psi(x_1,x_2,t)=-\psi(x_2,x_1,t)$ and only symmetric orbital wave functions need to be considered. Here, $\chi_{\uparrow}(\sigma_i)$,is the spin wave function of electron $i=1,2$ assumed to be in the spin-up ($\lvert \uparrow\rangle$) state. 

Now suppose that we inject the spin product state $\lvert\uparrow\rangle_{S_1}\lvert \downarrow\rangle_{S_2}$ where $S_1$ and $S_2$ refer to the two injection sources (see Fig.~\ref{fig_1}). Injection of electrons with a definite spin direction could be achieved using the spin-filtering effects of quantum dots in the Coulomb blockade regime and subjected to a magnetic field \cite{Recher2000}. Initially, the two electrons are well separated and localized near $S_1$ and $S_2$, respectively. So the initial total wave function takes on the form
\begin{multline}
    \Psi_{\text{in}}(x_1\sigma_1,x_2\sigma_2)=\frac{1}{\sqrt{2}}\left(\phi_{S_1}(x_1)\chi_{\uparrow}(\sigma_1)\phi_{S_2}(x_2)\chi_{\downarrow}(\sigma_2)-\right.\\
    \left.\phi_{S_1}(x_2)\chi_{\uparrow}(\sigma_2)\phi_{S_2}(x_1)\chi_{\downarrow}(\sigma_1)\right).
\end{multline}
This wave function can be more conveniently written in second quantized form \\ $\Psi_{\text{in}}(x_1\sigma_1,x_2\sigma_2)=\langle x_1\sigma_1,x_2\sigma_2|d_{S_1\uparrow}^{\dagger}d_{S_2\downarrow}^{\dagger}|0\rangle$ where $d_{S_i\sigma}^{\dagger}$ creates an electron with spin $\sigma$ near source $S_i$, $i=1,2$ and $|0\rangle$ is the particle vacuum. The corresponding state ket $|\Psi_{\text{in}}\rangle=d_{S_1\uparrow}^{\dagger}d_{S_2\downarrow}^{\dagger}|0\rangle$ can be written as the sum of a $S=0$ spin singlet $|S\rangle=(1/\sqrt{2})(d_{S_1\uparrow}^{\dagger}d_{S_2\downarrow}^{\dagger}-d_{S_1\downarrow}^{\dagger}d_{S_2\uparrow}^{\dagger})|0\rangle$  and a $S=1,S_z=0$ spin triplet $|T_0\rangle=(1/\sqrt{2})(d_{S_1\uparrow}^{\dagger}d_{S_2\downarrow}^{\dagger}-d_{S_1\downarrow}^{\dagger} d_{S_2\uparrow}^{\dagger})|0\rangle$ as 
\begin{equation}
|\Psi_{\text{in}}\rangle=\frac{1}{\sqrt{2}}(|S\rangle+|T_0\rangle)
\end{equation}
To make contact with our discussion about the form of the wave function, we note that 
\begin{multline}
\langle x_1\sigma_1,x_2\sigma_2|S,T_0\rangle=\frac{1}{\sqrt{2}}\left(\chi_{\uparrow}(\sigma_1)\chi_{\downarrow}(\sigma_2)\mp\chi_{\uparrow}(\sigma_2)\chi_{\downarrow}(\sigma_1)\right)\\
\times\frac{1}{\sqrt{2}}\left(\phi_{S_1}(x_1)\phi_{S_2}(x_2)\pm\phi_{S_1}(x_2)\phi_{S_2}(x_1)\right) \, ,
\end{multline}
with upper (lower) sign for $S$ ($T_0$).
As ${\bm S}^2$ and $S_z$ commute with $H$, the singlet and triplet parts of the wave function $\Psi_{\text{in}}(x_1\sigma_1,x_2\sigma_2)$ are conserved during the time evolution and when approaching each other at the beamsplitter. As the orbital part of the wave function will separate in a center of mass and relative coordinate part at long times we have 
$\psi(x_1,x_2,t)=\psi_{\rm r}(x,t)\,\psi_{\rm cm}(X,t)$ (cf. Eq.~(\ref{psi_sep})) with $\psi_{\rm r}^{S,T_0}(x,t)=\pm \psi_{\rm r}^{S,T_0}(-x,t)$ for the singlet part ($+$ sign) and triplet part ($-$ sign). Since time evolution is governed by a linear operator $|\Psi_{\text{out}}(t)\rangle=\exp(-iHt/\hbar)|\Psi_{\text{in}}\rangle$, we have 
\begin{equation}
    |\Psi_{\text{out}}(t)\rangle=\frac{1}{\sqrt{2}}\left(|S(t)\rangle+|T_{0}(t)\rangle\right).
\end{equation}
The probability density that two electrons are detected in a singlet state at corresponding detectors $D_1$ and $D_2$ during a time interval $\Delta t$ around time $t$ is 
\begin{multline}
P_D^{S}(t,\Delta t)=\frac{1}{2}\int_{t-\Delta t/2}^{t+\Delta t/2}dt'\, \langle S(t')|
\bigl (|x_{D1},x_{D2}\rangle\langle x_{D1},x_{D2}|\\
+|x_{D2},x_{D1}\rangle\langle x_{D2},x_{D1}|\bigr)|S(t')\rangle \\
=\int_{t-\Delta t/2}^{t+\Delta t/2}\,dt'|\psi_r^{S}(|x_{D_1}-x_{D_2}|,t')|^2\\
\times|\psi_{\rm cm}^{S}((x_{D_1}+x_{D_2})/2,t')|^2,
\end{multline}
and for $P_D^{T_0}(t,\Delta t)$ correspondingly. Note that at long times $t$, as discussed in Subsections \ref{sec:ModelHamiltonian} 3. and 4., the probability to stay close to the beamsplitter region is ruled by quantum tunneling into the anti-bound states. Even though the survival probability is diminished for the center of mass coordinate being still at the center by an exponential decay $~e^{-\lambda t}$ (see Fig.~\ref{fig_6}), the relative wave function undergoes quantum tunneling through anti-bound states with different lifetimes for singlet and triplet states (see Fig.~\ref{fig_8}). For two electrons observed at different detectors $D_1$ and $D_2$, the relative coordinate needs to have tunneled out of the molecule. For the specific example shown in Fig.~\ref{fig_8}, this is much more likely for the singlet as for the triplet in the time range where the most coupled state dominates the decay ($t=30\ldots40$). Therefore, detection of delocalized entanglement becomes possible.

Coincident late arrival of electrons to two detectors selects for a spin-entangled state such that a choice between a singlet or a triplet can be tuned by tuning the beamsplitter potential and choosing the appropriate time delay for detection. On a related note, tunable loading of stable singlets or triplets in a quantum dot based on tunneling rate separation~\cite{Wenz2019} and coherent transport of entangled electron spin pairs over several $\rm{\mu m}$~\cite{Jadot2021} have been recently demonstrated. Hence the two-electron collider can serve as stochastic source of mobile EPR pairs, i.e.\ the two electrons are separated in (orbital) space to two distant detectors but still correlated in the spin degree of freedom. The detection of such spin-correlations could be achieved via a Bell test using tunable spin filters combined with charge noise measurements \cite{Kawabata2001, Chtchelkatchev2002, Recher2004}.

{\section{Conclusions\label{secConclusion}}
We have analyzed the problem of two interacting electrons scattering at a saddle point potential in the quantum Hall regime. By introducing canonically conjugate guiding centers for the drift motion in two dimensions the problem reduces to that of two electrons interacting via strong long-range Coulomb interaction and subjected to the saddle point potential. We have shown that the problem remains exactly separable in relative  and center-of-mass coordinates. These coordinates are well suited to describe effects of anti-bound states of electron pairs in the scattering problem of Hanbury Brown and Twiss and Hong-Ou-Mandel setups of electron optics. Our focus lies on the study of 
critical trajectories where two injected electrons stay for a long time at the beamsplitter. We found characteristic long-time tails for the corresponding probabilities using classical phase space arguments as well as quantum mechanical effects of the two-particle problem. This leads to the information about the beamspitter potential, the nature of the interaction between two electrons as well as to the characteristic tendency to form quasi-bound states. The latter effect is shown to be different for symmetric and antisymmetric orbital wave functions distinguishing spin-singlets from spin-triplets. Our results may motivate experiments in the field of two-particle quantum optics with a time-resolved detection scheme.

\begin{acknowledgments}
We are thankful for insightful discussions with P.~W.~Brouwer. PS and PR acknowledge support from the Deutsche Forschungsgemeinschaft (DFG, German Research Foundation) under Germany’s Excellence Strategy -- EXC-2123 QuantumFrontiers -- 390837967. VK has been supported by grant no.~lzp-2021/1-0232 from the Latvian Council of Science~and the Latvian Quantum Initiative within European Union Recovery and Resilience Facility project no.~2.3.1.1.i.0/1/22/I/CFLA/001.
\end{acknowledgments}

\appendix*
\section{Analytical derivations}

{\it Trajectories asymptotics.} Before considering the specific critical examples, like Eqs.~(\ref{eps12}) and (\ref{eq:energytransfer}), 
we need to describe 
the asymptotics of generic trajectories 
in the HOM setup at the parabolic beamsplitter.

Individual electron's trajectories are found straightforwardly from
the relative coordinate and the center of mass trajectories. 
The generic relative coordinate trajectory described by the effective Hamiltonian $H^{\rm eff}_{\rm r}$, Eqs.~(\ref{H_r}) may be chosen to be symmetric in time with the asymptotics 
 \begin{align}\label{alpha_beta_r}
&x(\lambda|t|\gg 1)=-\fr{\alpha_{\rm r}}{\sqrt{A}}e^{\lambda |t|}
-\fr{\beta_{\rm r}}{\sqrt{A}}e^{-\lambda |t|} \ , 
 \\
&
 y(\lambda|t|\gg 1)=-\fr{t}{|t|}\fr{\alpha_{\rm r}}{\sqrt{C}}e^{\lambda |t|} 
 + \fr{t}{|t|}\fr{\beta_{\rm r}}{\sqrt{C}}e^{-\lambda |t|}\ , \nonumber
 \end{align}
with arbitrary coefficients $\alpha_{\rm r}, \beta_{\rm r}$. At very large $\lambda|t|$ (either for negative or positive times $t$) the first term
$\sim\alpha_{\rm r}$ in Eqs.~(\ref{alpha_beta_r}), describing the drift along the separatrix, dominates.
The second $\sim \beta_{\rm r}$ term in Eqs.~(\ref{alpha_beta_r}) describes the asymptotic squeezing of the trajectories around the separatrix.
Taking into account these terms would allow us to find the electrons' energies before and after collision.

Together with the proper redefinition of the zero of time, $t=0$, Eqs.~(\ref{alpha_beta_r}) describe all the relative coordinate trajectories except two. 
The two remaining critical trajectories are those where one starts with a very large displacement (either at the upper left, or the lower right corners of the $(x,y)$ plane in our example), 
but then the two electrons stay almost frozen, asymptotically approaching one of the stationary points $(x,y)=(0,\pm 2y_0)$. 
Knowing the trajectories with the inter-electron distance frozen at $t\rightarrow\infty$ is necessary 
for deriving Eq.~(\ref{eps12}).
The negative time asymptotics of these critical trajectories is still given by Eqs.~(\ref{alpha_beta_r}) with the product of the coefficients 
$\alpha_{\rm r}\beta_{\rm r}$ having a unique value (to be found below).


The center of mass classical trajectory described by the quadratic in coordinates effective Hamiltonian 
$H^{\rm eff}_{\rm cm}$, Eq.~(\ref{H_cm}), may be found exactly 
 \begin{align}\label{alpha_beta_CM}
&X(t)=\fr{\alpha_{\rm cm}}{\sqrt{A}}e^{-\lambda t}
-\fr{\beta_{\rm cm}}{\sqrt{A}}e^{\lambda t} \ , 
 \\
& 
 Y(t)=-\fr{\alpha_{\rm cm}}{\sqrt{C}}e^{-\lambda t} 
 - \fr{\beta_{\rm cm}}{\sqrt{C}}e^{\lambda t}\ .\nonumber
 \end{align}
Here again $\alpha_{\rm cm}, \beta_{\rm cm}$ are two arbitrary coefficients. 

Next, using that the classical single-electron energies are $\eps_i=-Ax_i^2/2 +Cy_i^2/2$, $i=1,2$, 
with the help of Eqs.~(\ref{alpha_beta_r}, \ref{alpha_beta_CM})
we find the energies of electrons before the collision, at negative times $\lambda t\ll -1$
 \begin{align}\label{eps_before}
\eps_{1(2)\, {\rm in}}=\alpha_{\rm cm}\beta_{\rm cm} \pm [\alpha_{\rm cm}\beta_{\rm r} -\beta_{\rm cm}\alpha_{\rm r}] -\alpha_{\rm r}\beta_{\rm r}/4 \ , 
 \end{align}
and after the collision, at  $\lambda t\gg 1$, 
 \begin{align}\label{A.4}
\eps_{1(2)\, {\rm out}}=\alpha_{\rm cm}\beta_{\rm cm} \pm [\alpha_{\rm cm}\alpha_{\rm r} -\beta_{\rm cm}\beta_{\rm r}] -\alpha_{\rm r}\beta_{\rm r}/4 \ . 
 \end{align}
Here in both cases in/out the upper sign $(+)$ is for  electron 1 and the lower sign $(-)$ for  electron 2.

The time shifts between the ingoing and outgoing electrons, $\Delta t_{\rm in}$ and $\Delta t_{\rm out}$, are found as
 \begin{align}
x_1(t)/x_2(t)&\underset{t\rightarrow -\infty}{=}
-e^{-\lambda\Delta t_{\rm in}} \ , \nonumber \\
x_1(t)/x_2(t)&\underset{t\rightarrow +\infty}{=}
-e^{\lambda\Delta t_{\rm out}} \ . 
 \end{align}
Using $y_{1,2}$ instead of $x_{1,2}$ would obviously be the same. 
From Eqs.~(\ref{alpha_beta_r}, \ref{alpha_beta_CM}) we deduce
 \bq\label{Delta_t_out}
e^{\lambda\Delta t_{\rm in}} =\fr{\alpha_{\rm
r}+2\alpha_{\rm cm}}{\alpha_{\rm
r}-2\alpha_{\rm cm}} \ ,
\ e^{\lambda\Delta t_{\rm out}} =\fr{\alpha_{\rm
r}+2\beta_{\rm cm}}{\alpha_{\rm
r}-2\beta_{\rm cm}} \ .
 \ee

{\it Derivation of Eq.~(\ref{eps12}):} 
We are searching for a special solution of the equations of motion where two electrons with the initial energies $\eps_{1 \rm in}, \eps_{2 \rm in}$ enter the beamsplitter 
with the time delay $\Delta t_{\rm in}$ and stay there forever, approaching asymptotically the stationary positions $(x,y)=(0,\pm 2y_0)$.
This means choosing the solution Eqs.~(\ref{alpha_beta_CM}) with $\beta_{\rm cm}=0$.
The coefficient $\alpha_{\rm cm}$ remains undetermined.
Varying $\alpha_{\rm cm}$ leads to the time shift of the trajectory,
$\alpha_{\rm cm}>0$ means
the electron $1$ coming from the left 
enters the beamsplitter first.

The relation between the coefficients $\alpha_{\rm r}$ and $\beta_{\rm r}$ for the critical trajectory with stationary relative coordinate after the collision is found from its energy $-Ax^2/4 +Cy^2/4=2\eps_0$ (see $H^{\rm eff}_{\rm r}$, Eq.~(\ref{H_r})), yielding
 \bq
\beta_{\rm r}\alpha_{\rm r} =-2\eps_0 \ . 
 \ee
Finally, the first of Eqs.~(\ref{Delta_t_out}) may be written as 
 \bq\label{A.8}
\alpha_{\rm cm} =\tanh(\lambda\Delta t_{\rm in}/2)\fr{\alpha_{\rm r}}{2} \ .
 \ee
Substituting the last two results into Eq.~(\ref{eps_before}) yields Eq.~(\ref{eps12}).

{\it Towards derivation of Eq.~(\ref{eq:energytransfer}):} 
Now we want to consider the case of injection of two electrons with non-interacting half transmission energies  
$\eps_{1 \rm in}= \eps_{2 \rm in}=0$ but with a finite time delay $\Delta t_{\rm in}$.
Mathematically that means $\beta_{\rm r}= \beta_{\rm cm}=0$ and arbitrary
$\alpha_{\rm cm}$. The remaining coefficient may be written in a form 
 \bq\label{A.9}
\alpha_{\rm r}= \sqrt{2\gamma_0\eps_0} \ .
 \ee 
The purely numerical coefficient $\gamma_0\sim 1$ here depending on the ratio $A/C$,
could not be found with the simple considerations presented here.
Nevertheless, it may be extracted from the exact solutions of equations of motion developed in \cite{Pavlovska2022}, $\gamma_0=2[2A/(A+C)]^{1/3}/3$. 
Combining Eqs.~(\ref{A.4},\ref{A.8},\ref{A.9}) yields Eq.~(\ref{eq:energytransfer}).


\begin{thebibliography}{49}%
\makeatletter
\providecommand \@ifxundefined [1]{%
 \@ifx{#1\undefined}
}%
\providecommand \@ifnum [1]{%
 \ifnum #1\expandafter \@firstoftwo
 \else \expandafter \@secondoftwo
 \fi
}%
\providecommand \@ifx [1]{%
 \ifx #1\expandafter \@firstoftwo
 \else \expandafter \@secondoftwo
 \fi
}%
\providecommand \natexlab [1]{#1}%
\providecommand \enquote  [1]{``#1''}%
\providecommand \bibnamefont  [1]{#1}%
\providecommand \bibfnamefont [1]{#1}%
\providecommand \citenamefont [1]{#1}%
\providecommand \href@noop [0]{\@secondoftwo}%
\providecommand \href [0]{\begingroup \@sanitize@url \@href}%
\providecommand \@href[1]{\@@startlink{#1}\@@href}%
\providecommand \@@href[1]{\endgroup#1\@@endlink}%
\providecommand \@sanitize@url [0]{\catcode `\\12\catcode `\$12\catcode
  `\&12\catcode `\#12\catcode `\^12\catcode `\_12\catcode `\%12\relax}%
\providecommand \@@startlink[1]{}%
\providecommand \@@endlink[0]{}%
\providecommand \url  [0]{\begingroup\@sanitize@url \@url }%
\providecommand \@url [1]{\endgroup\@href {#1}{\urlprefix }}%
\providecommand \urlprefix  [0]{URL }%
\providecommand \Eprint [0]{\href }%
\providecommand \doibase [0]{https://doi.org/}%
\providecommand \selectlanguage [0]{\@gobble}%
\providecommand \bibinfo  [0]{\@secondoftwo}%
\providecommand \bibfield  [0]{\@secondoftwo}%
\providecommand \translation [1]{[#1]}%
\providecommand \BibitemOpen [0]{}%
\providecommand \bibitemStop [0]{}%
\providecommand \bibitemNoStop [0]{.\EOS\space}%
\providecommand \EOS [0]{\spacefactor3000\relax}%
\providecommand \BibitemShut  [1]{\csname bibitem#1\endcsname}%
\let\auto@bib@innerbib\@empty
\bibitem [{\citenamefont {Bocquillon}\ \emph {et~al.}(2014)\citenamefont
  {Bocquillon}, \citenamefont {Freulon}, \citenamefont {Parmentier},
  \citenamefont {Berroir}, \citenamefont {Pla{\c{c}}ais}, \citenamefont {Wahl},
  \citenamefont {Rech}, \citenamefont {Jonckheere}, \citenamefont {Martin},
  \citenamefont {Grenier}, \citenamefont {Ferraro}, \citenamefont
  {Degiovanni},\ and\ \citenamefont {F{\`{e}}ve}}]{Bocquillon2014}%
  \BibitemOpen
  \bibfield  {author} {\bibinfo {author} {\bibfnamefont {E.}~\bibnamefont
  {Bocquillon}}, \bibinfo {author} {\bibfnamefont {V.}~\bibnamefont {Freulon}},
  \bibinfo {author} {\bibfnamefont {F.~D.}\ \bibnamefont {Parmentier}},
  \bibinfo {author} {\bibfnamefont {J.-M.}\ \bibnamefont {Berroir}}, \bibinfo
  {author} {\bibfnamefont {B.}~\bibnamefont {Pla{\c{c}}ais}}, \bibinfo {author}
  {\bibfnamefont {C.}~\bibnamefont {Wahl}}, \bibinfo {author} {\bibfnamefont
  {J.}~\bibnamefont {Rech}}, \bibinfo {author} {\bibfnamefont {T.}~\bibnamefont
  {Jonckheere}}, \bibinfo {author} {\bibfnamefont {T.}~\bibnamefont {Martin}},
  \bibinfo {author} {\bibfnamefont {C.}~\bibnamefont {Grenier}}, \bibinfo
  {author} {\bibfnamefont {D.}~\bibnamefont {Ferraro}}, \bibinfo {author}
  {\bibfnamefont {P.}~\bibnamefont {Degiovanni}},\ and\ \bibinfo {author}
  {\bibfnamefont {G.}~\bibnamefont {F{\`{e}}ve}},\ }\bibfield  {title}
  {\bibinfo {title} {{Electron quantum optics in ballistic chiral
  conductors}},\ }\href {https://doi.org/10.1002/andp.201300181} {\bibfield
  {journal} {\bibinfo  {journal} {Annalen der Physik}\ }\textbf {\bibinfo
  {volume} {526}},\ \bibinfo {pages} {1} (\bibinfo {year} {2014})}\BibitemShut
  {NoStop}%
\bibitem [{\citenamefont {B{\"{a}}uerle}\ \emph {et~al.}(2018)\citenamefont
  {B{\"{a}}uerle}, \citenamefont {Christian~Glattli}, \citenamefont {Meunier},
  \citenamefont {Portier}, \citenamefont {Roche}, \citenamefont {Roulleau},
  \citenamefont {Takada},\ and\ \citenamefont {Waintal}}]{Bauerle2018}%
  \BibitemOpen
  \bibfield  {author} {\bibinfo {author} {\bibfnamefont {C.}~\bibnamefont
  {B{\"{a}}uerle}}, \bibinfo {author} {\bibfnamefont {D.}~\bibnamefont
  {Christian~Glattli}}, \bibinfo {author} {\bibfnamefont {T.}~\bibnamefont
  {Meunier}}, \bibinfo {author} {\bibfnamefont {F.}~\bibnamefont {Portier}},
  \bibinfo {author} {\bibfnamefont {P.}~\bibnamefont {Roche}}, \bibinfo
  {author} {\bibfnamefont {P.}~\bibnamefont {Roulleau}}, \bibinfo {author}
  {\bibfnamefont {S.}~\bibnamefont {Takada}},\ and\ \bibinfo {author}
  {\bibfnamefont {X.}~\bibnamefont {Waintal}},\ }\bibfield  {title} {\bibinfo
  {title} {{Coherent control of single electrons: a review of current
  progress}},\ }\href {https://doi.org/10.1088/1361-6633/aaa98a} {\bibfield
  {journal} {\bibinfo  {journal} {Reports on Progress in Physics}\ }\textbf
  {\bibinfo {volume} {81}},\ \bibinfo {pages} {056503} (\bibinfo {year}
  {2018})}\BibitemShut {NoStop}%
\bibitem [{\citenamefont {Henny}\ \emph {et~al.}(1999)\citenamefont {Henny},
  \citenamefont {Oberholzer}, \citenamefont {Strunk}, \citenamefont {Heinzel},
  \citenamefont {Ensslin}, \citenamefont {Holland},\ and\ \citenamefont
  {Sch\"{o}nenberger}}]{Henny1999}%
  \BibitemOpen
  \bibfield  {author} {\bibinfo {author} {\bibfnamefont {M.}~\bibnamefont
  {Henny}}, \bibinfo {author} {\bibfnamefont {S.}~\bibnamefont {Oberholzer}},
  \bibinfo {author} {\bibfnamefont {C.}~\bibnamefont {Strunk}}, \bibinfo
  {author} {\bibfnamefont {T.}~\bibnamefont {Heinzel}}, \bibinfo {author}
  {\bibfnamefont {K.}~\bibnamefont {Ensslin}}, \bibinfo {author} {\bibfnamefont
  {M.}~\bibnamefont {Holland}},\ and\ \bibinfo {author} {\bibfnamefont
  {C.}~\bibnamefont {Sch\"{o}nenberger}},\ }\bibfield  {title} {\bibinfo
  {title} {{The Fermionic Hanbury Brown and Twiss Experiment}},\ }\href
  {https://doi.org/10.1126/science.284.5412.296} {\bibfield  {journal}
  {\bibinfo  {journal} {Science}\ }\textbf {\bibinfo {volume} {284}},\ \bibinfo
  {pages} {296} (\bibinfo {year} {1999})}\BibitemShut {NoStop}%
\bibitem [{\citenamefont {Oliver}\ \emph {et~al.}(1999)\citenamefont {Oliver},
  \citenamefont {Kim}, \citenamefont {Liu},\ and\ \citenamefont
  {Yamamoto}}]{Oliver1999}%
  \BibitemOpen
  \bibfield  {author} {\bibinfo {author} {\bibfnamefont {W.~D.}\ \bibnamefont
  {Oliver}}, \bibinfo {author} {\bibfnamefont {J.}~\bibnamefont {Kim}},
  \bibinfo {author} {\bibfnamefont {R.~C.}\ \bibnamefont {Liu}},\ and\ \bibinfo
  {author} {\bibfnamefont {Y.}~\bibnamefont {Yamamoto}},\ }\bibfield  {title}
  {\bibinfo {title} {{Hanbury Brown and Twiss-Type Experiment with
  Electrons}},\ }\href {https://doi.org/10.1126/science.284.5412.299}
  {\bibfield  {journal} {\bibinfo  {journal} {Science}\ }\textbf {\bibinfo
  {volume} {284}},\ \bibinfo {pages} {299} (\bibinfo {year}
  {1999})}\BibitemShut {NoStop}%
\bibitem [{\citenamefont {F{\`{e}}ve}\ \emph {et~al.}(2007)\citenamefont
  {F{\`{e}}ve}, \citenamefont {Mah{\'{e}}}, \citenamefont {Berroir},
  \citenamefont {Kontos}, \citenamefont {Pla{\c{c}}ais}, \citenamefont
  {Glattli}, \citenamefont {Cavanna}, \citenamefont {Etienne}, \citenamefont
  {Jin}, \citenamefont {Feve}, \citenamefont {Mahe}, \citenamefont {Berroir},
  \citenamefont {Kontos}, \citenamefont {Placais}, \citenamefont {Glattli},
  \citenamefont {Cavanna}, \citenamefont {Etienne},\ and\ \citenamefont
  {Jin}}]{Feve2007}%
  \BibitemOpen
  \bibfield  {author} {\bibinfo {author} {\bibfnamefont {G.}~\bibnamefont
  {F{\`{e}}ve}}, \bibinfo {author} {\bibfnamefont {A.}~\bibnamefont
  {Mah{\'{e}}}}, \bibinfo {author} {\bibfnamefont {J.-M.}\ \bibnamefont
  {Berroir}}, \bibinfo {author} {\bibfnamefont {T.}~\bibnamefont {Kontos}},
  \bibinfo {author} {\bibfnamefont {B.}~\bibnamefont {Pla{\c{c}}ais}}, \bibinfo
  {author} {\bibfnamefont {D.~C.}\ \bibnamefont {Glattli}}, \bibinfo {author}
  {\bibfnamefont {A.}~\bibnamefont {Cavanna}}, \bibinfo {author} {\bibfnamefont
  {B.}~\bibnamefont {Etienne}}, \bibinfo {author} {\bibfnamefont
  {Y.}~\bibnamefont {Jin}}, \bibinfo {author} {\bibfnamefont {G.}~\bibnamefont
  {Feve}}, \bibinfo {author} {\bibfnamefont {A.}~\bibnamefont {Mahe}}, \bibinfo
  {author} {\bibfnamefont {J.-M.}\ \bibnamefont {Berroir}}, \bibinfo {author}
  {\bibfnamefont {T.}~\bibnamefont {Kontos}}, \bibinfo {author} {\bibfnamefont
  {B.}~\bibnamefont {Placais}}, \bibinfo {author} {\bibfnamefont {D.~C.}\
  \bibnamefont {Glattli}}, \bibinfo {author} {\bibfnamefont {A.}~\bibnamefont
  {Cavanna}}, \bibinfo {author} {\bibfnamefont {B.}~\bibnamefont {Etienne}},\
  and\ \bibinfo {author} {\bibfnamefont {Y.}~\bibnamefont {Jin}},\ }\bibfield
  {title} {\bibinfo {title} {{An on-demand coherent single-electron source.}},\
  }\href {https://doi.org/10.1126/science.1141243} {\bibfield  {journal}
  {\bibinfo  {journal} {Science}\ }\textbf {\bibinfo {volume}
  {316}},\ \bibinfo {pages} {1169} (\bibinfo {year} {2007})}\BibitemShut
  {NoStop}%
\bibitem [{\citenamefont {Bocquillon}\ \emph {et~al.}(2012)\citenamefont
  {Bocquillon}, \citenamefont {Parmentier}, \citenamefont {Grenier},
  \citenamefont {Berroir}, \citenamefont {Degiovanni}, \citenamefont {Glattli},
  \citenamefont {Pla{\c{c}}ais}, \citenamefont {Cavanna}, \citenamefont {Jin},\
  and\ \citenamefont {F{\`{e}}ve}}]{Feve2012exp}%
  \BibitemOpen
  \bibfield  {author} {\bibinfo {author} {\bibfnamefont {E.}~\bibnamefont
  {Bocquillon}}, \bibinfo {author} {\bibfnamefont {F.~D.}\ \bibnamefont
  {Parmentier}}, \bibinfo {author} {\bibfnamefont {C.}~\bibnamefont {Grenier}},
  \bibinfo {author} {\bibfnamefont {J.-M.}\ \bibnamefont {Berroir}}, \bibinfo
  {author} {\bibfnamefont {P.}~\bibnamefont {Degiovanni}}, \bibinfo {author}
  {\bibfnamefont {D.~C.}\ \bibnamefont {Glattli}}, \bibinfo {author}
  {\bibfnamefont {B.}~\bibnamefont {Pla{\c{c}}ais}}, \bibinfo {author}
  {\bibfnamefont {A.}~\bibnamefont {Cavanna}}, \bibinfo {author} {\bibfnamefont
  {Y.}~\bibnamefont {Jin}},\ and\ \bibinfo {author} {\bibfnamefont
  {G.}~\bibnamefont {F{\`{e}}ve}},\ }\bibfield  {title} {\bibinfo {title}
  {{Electron Quantum Optics: Partitioning Electrons One by One}},\ }\href
  {https://doi.org/10.1103/PhysRevLett.108.196803} {\bibfield  {journal}
  {\bibinfo  {journal} {Physical Review Letters}\ }\textbf {\bibinfo {volume}
  {108}},\ \bibinfo {pages} {196803} (\bibinfo {year} {2012})}\BibitemShut
  {NoStop}%
\bibitem [{\citenamefont {Bocquillon}\ \emph {et~al.}(2013)\citenamefont
  {Bocquillon}, \citenamefont {Freulon}, \citenamefont {Berroir}, \citenamefont
  {Degiovanni}, \citenamefont {Pla{\c{c}}ais}, \citenamefont {Cavanna},
  \citenamefont {Jin},\ and\ \citenamefont {F{\`{e}}ve}}]{Bocquillon2013}%
  \BibitemOpen
  \bibfield  {author} {\bibinfo {author} {\bibfnamefont {E.}~\bibnamefont
  {Bocquillon}}, \bibinfo {author} {\bibfnamefont {V.}~\bibnamefont {Freulon}},
  \bibinfo {author} {\bibfnamefont {J.-M.}\ \bibnamefont {Berroir}}, \bibinfo
  {author} {\bibfnamefont {P.}~\bibnamefont {Degiovanni}}, \bibinfo {author}
  {\bibfnamefont {B.}~\bibnamefont {Pla{\c{c}}ais}}, \bibinfo {author}
  {\bibfnamefont {A.}~\bibnamefont {Cavanna}}, \bibinfo {author} {\bibfnamefont
  {Y.}~\bibnamefont {Jin}},\ and\ \bibinfo {author} {\bibfnamefont
  {G.}~\bibnamefont {F{\`{e}}ve}},\ }\bibfield  {title} {\bibinfo {title}
  {{Coherence and indistinguishability of single electrons emitted by
  independent sources.}},\ }\href {https://doi.org/10.1126/science.1232572}
  {\bibfield  {journal} {\bibinfo  {journal} {Science}\
  }\textbf {\bibinfo {volume} {339}},\ \bibinfo {pages} {1054} (\bibinfo {year}
  {2013})}\BibitemShut {NoStop}%
\bibitem [{\citenamefont {Jullien}\ \emph {et~al.}(2014)\citenamefont
  {Jullien}, \citenamefont {Roulleau}, \citenamefont {Roche}, \citenamefont
  {Cavanna}, \citenamefont {Jin},\ and\ \citenamefont {Glattli}}]{Jullien2014}%
  \BibitemOpen
  \bibfield  {author} {\bibinfo {author} {\bibfnamefont {T.}~\bibnamefont
  {Jullien}}, \bibinfo {author} {\bibfnamefont {P.}~\bibnamefont {Roulleau}},
  \bibinfo {author} {\bibfnamefont {B.}~\bibnamefont {Roche}}, \bibinfo
  {author} {\bibfnamefont {A.}~\bibnamefont {Cavanna}}, \bibinfo {author}
  {\bibfnamefont {Y.}~\bibnamefont {Jin}},\ and\ \bibinfo {author}
  {\bibfnamefont {D.~C.}\ \bibnamefont {Glattli}},\ }\bibfield  {title}
  {\bibinfo {title} {{Quantum tomography of an electron.}},\ }\href
  {https://doi.org/10.1038/nature13821} {\bibfield  {journal} {\bibinfo
  {journal} {Nature}\ }\textbf {\bibinfo {volume} {514}},\ \bibinfo {pages}
  {603} (\bibinfo {year} {2014})}\BibitemShut {NoStop}%
\bibitem [{\citenamefont {Bisognin}\ \emph {et~al.}(2019)\citenamefont
  {Bisognin}, \citenamefont {Marguerite}, \citenamefont {Roussel},
  \citenamefont {Kumar}, \citenamefont {Cabart}, \citenamefont {Chapdelaine},
  \citenamefont {Mohammad-Djafari}, \citenamefont {Berroir}, \citenamefont
  {Bocquillon}, \citenamefont {Pla{\c{c}}ais}, \citenamefont {Cavanna},
  \citenamefont {Gennser}, \citenamefont {Jin}, \citenamefont {Degiovanni},\
  and\ \citenamefont {F{\`{e}}ve}}]{Bisognin2019}%
  \BibitemOpen
  \bibfield  {author} {\bibinfo {author} {\bibfnamefont {R.}~\bibnamefont
  {Bisognin}}, \bibinfo {author} {\bibfnamefont {A.}~\bibnamefont
  {Marguerite}}, \bibinfo {author} {\bibfnamefont {B.}~\bibnamefont {Roussel}},
  \bibinfo {author} {\bibfnamefont {M.}~\bibnamefont {Kumar}}, \bibinfo
  {author} {\bibfnamefont {C.}~\bibnamefont {Cabart}}, \bibinfo {author}
  {\bibfnamefont {C.}~\bibnamefont {Chapdelaine}}, \bibinfo {author}
  {\bibfnamefont {A.}~\bibnamefont {Mohammad-Djafari}}, \bibinfo {author}
  {\bibfnamefont {J.-M.}\ \bibnamefont {Berroir}}, \bibinfo {author}
  {\bibfnamefont {E.}~\bibnamefont {Bocquillon}}, \bibinfo {author}
  {\bibfnamefont {B.}~\bibnamefont {Pla{\c{c}}ais}}, \bibinfo {author}
  {\bibfnamefont {A.}~\bibnamefont {Cavanna}}, \bibinfo {author} {\bibfnamefont
  {U.}~\bibnamefont {Gennser}}, \bibinfo {author} {\bibfnamefont
  {Y.}~\bibnamefont {Jin}}, \bibinfo {author} {\bibfnamefont {P.}~\bibnamefont
  {Degiovanni}},\ and\ \bibinfo {author} {\bibfnamefont {G.}~\bibnamefont
  {F{\`{e}}ve}},\ }\bibfield  {title} {\bibinfo {title} {{Quantum tomography of
  electrical currents}},\ }\href {https://doi.org/10.1038/s41467-019-11369-5}
  {\bibfield  {journal} {\bibinfo  {journal} {Nature Communications}\ }\textbf
  {\bibinfo {volume} {10}},\ \bibinfo {pages} {3379} (\bibinfo {year}
  {2019})}\BibitemShut {NoStop}%
\bibitem [{\citenamefont {Roussel}\ \emph {et~al.}(2021)\citenamefont
  {Roussel}, \citenamefont {Cabart}, \citenamefont {F{\`{e}}ve},\ and\
  \citenamefont {Degiovanni}}]{Rouseel-2021}%
  \BibitemOpen
  \bibfield  {author} {\bibinfo {author} {\bibfnamefont {B.}~\bibnamefont
  {Roussel}}, \bibinfo {author} {\bibfnamefont {C.}~\bibnamefont {Cabart}},
  \bibinfo {author} {\bibfnamefont {G.}~\bibnamefont {F{\`{e}}ve}},\ and\
  \bibinfo {author} {\bibfnamefont {P.}~\bibnamefont {Degiovanni}},\ }\bibfield
   {title} {\bibinfo {title} {{Processing Quantum Signals Carried by Electrical
  Currents}},\ }\href {https://doi.org/10.1103/PRXQuantum.2.020314} {\bibfield
  {journal} {\bibinfo  {journal} {PRX Quantum}\ }\textbf {\bibinfo {volume}
  {2}},\ \bibinfo {pages} {020314} (\bibinfo {year} {2021})}\BibitemShut
  {NoStop}%
\bibitem [{\citenamefont {Oliver}\ \emph {et~al.}(2002)\citenamefont {Oliver},
  \citenamefont {Yamaguchi},\ and\ \citenamefont {Yamamoto}}]{Oliver2002}%
  \BibitemOpen
  \bibfield  {author} {\bibinfo {author} {\bibfnamefont {W.~D.}\ \bibnamefont
  {Oliver}}, \bibinfo {author} {\bibfnamefont {F.}~\bibnamefont {Yamaguchi}},\
  and\ \bibinfo {author} {\bibfnamefont {Y.}~\bibnamefont {Yamamoto}},\
  }\bibfield  {title} {\bibinfo {title} {Electron entanglement via a quantum
  dot},\ }\href {https://doi.org/10.1103/PhysRevLett.88.037901} {\bibfield
  {journal} {\bibinfo  {journal} {Physical  Review Letters}\ }\textbf {\bibinfo
  {volume} {88}},\ \bibinfo {pages} {037901} (\bibinfo {year}
  {2002})}\BibitemShut {NoStop}%
\bibitem [{\citenamefont {Saraga}\ \emph {et~al.}(2004)\citenamefont {Saraga},
  \citenamefont {Altshuler}, \citenamefont {Loss},\ and\ \citenamefont
  {Westervelt}}]{Saraga2004}%
  \BibitemOpen
  \bibfield  {author} {\bibinfo {author} {\bibfnamefont {D.~S.}\ \bibnamefont
  {Saraga}}, \bibinfo {author} {\bibfnamefont {B.~L.}\ \bibnamefont
  {Altshuler}}, \bibinfo {author} {\bibfnamefont {D.}~\bibnamefont {Loss}},\
  and\ \bibinfo {author} {\bibfnamefont {R.~M.}\ \bibnamefont {Westervelt}},\
  }\bibfield  {title} {\bibinfo {title} {Coulomb scattering in a 2d interacting
  electron gas and production of epr pairs},\ }\href
  {https://doi.org/10.1103/PhysRevLett.92.246803} {\bibfield  {journal}
  {\bibinfo  {journal} {Physical Review Letters}\ }\textbf {\bibinfo {volume} {92}},\
  \bibinfo {pages} {246803} (\bibinfo {year} {2004})}\BibitemShut {NoStop}%
\bibitem [{\citenamefont {Saraga}\ \emph {et~al.}(2005)\citenamefont {Saraga},
  \citenamefont {Altshuler}, \citenamefont {Loss},\ and\ \citenamefont
  {Westervelt}}]{Saraga2005}%
  \BibitemOpen
  \bibfield  {author} {\bibinfo {author} {\bibfnamefont {D.~S.}\ \bibnamefont
  {Saraga}}, \bibinfo {author} {\bibfnamefont {B.~L.}\ \bibnamefont
  {Altshuler}}, \bibinfo {author} {\bibfnamefont {D.}~\bibnamefont {Loss}},\
  and\ \bibinfo {author} {\bibfnamefont {R.~M.}\ \bibnamefont {Westervelt}},\
  }\bibfield  {title} {\bibinfo {title} {Coulomb scattering cross section in a
  two-dimensional electron gas and production of entangled electrons},\ }\href
  {https://doi.org/10.1103/PhysRevB.71.045338} {\bibfield  {journal} {\bibinfo
  {journal} {Physical Review B}\ }\textbf {\bibinfo {volume} {71}},\ \bibinfo
  {pages} {045338} (\bibinfo {year} {2005})}\BibitemShut {NoStop}%
\bibitem [{\citenamefont {Schroer}\ \emph {et~al.}(2014)\citenamefont
  {Schroer}, \citenamefont {Braunecker}, \citenamefont {Levy~Yeyati},\ and\
  \citenamefont {Recher}}]{Schroer2014}%
  \BibitemOpen
  \bibfield  {author} {\bibinfo {author} {\bibfnamefont {A.}~\bibnamefont
  {Schroer}}, \bibinfo {author} {\bibfnamefont {B.}~\bibnamefont {Braunecker}},
  \bibinfo {author} {\bibfnamefont {A.}~\bibnamefont {Levy~Yeyati}},\ and\
  \bibinfo {author} {\bibfnamefont {P.}~\bibnamefont {Recher}},\ }\bibfield
  {title} {\bibinfo {title} {Detection of spin entanglement via spin-charge
  separation in crossed tomonaga-luttinger liquids},\ }\href
  {https://doi.org/10.1103/PhysRevLett.113.266401} {\bibfield  {journal}
  {\bibinfo  {journal} {Physical Review Letters}\ }\textbf {\bibinfo {volume} {113}},\
  \bibinfo {pages} {266401} (\bibinfo {year} {2014})}\BibitemShut {NoStop}%
\bibitem [{\citenamefont {Ryu}\ and\ \citenamefont {Sim}(2022)}]{RyuSim2022b}%
  \BibitemOpen
  \bibfield  {author} {\bibinfo {author} {\bibfnamefont {S.}~\bibnamefont
  {Ryu}}\ and\ \bibinfo {author} {\bibfnamefont {H.-S.}\ \bibnamefont {Sim}},\
  }\bibfield  {title} {\bibinfo {title} {{Partition of Two Interacting
  Electrons by a Potential Barrier}},\ }\href
  {https://doi.org/10.1103/PhysRevLett.129.166801} {\bibfield  {journal}
  {\bibinfo  {journal} {Physical Review Letters}\ }\textbf {\bibinfo {volume}
  {129}},\ \bibinfo {pages} {166801} (\bibinfo {year} {2022})}\BibitemShut
  {NoStop}%
\bibitem [{\citenamefont {Leicht}\ \emph {et~al.}(2011)\citenamefont {Leicht},
  \citenamefont {Mirovsky}, \citenamefont {Kaestner}, \citenamefont {Hohls},
  \citenamefont {Kashcheyevs}, \citenamefont {Kurganova}, \citenamefont
  {Zeitler}, \citenamefont {Weimann}, \citenamefont {Pierz},\ and\
  \citenamefont {Schumacher}}]{Leicht2011}%
  \BibitemOpen
  \bibfield  {author} {\bibinfo {author} {\bibfnamefont {C.}~\bibnamefont
  {Leicht}}, \bibinfo {author} {\bibfnamefont {P.}~\bibnamefont {Mirovsky}},
  \bibinfo {author} {\bibfnamefont {B.}~\bibnamefont {Kaestner}}, \bibinfo
  {author} {\bibfnamefont {F.}~\bibnamefont {Hohls}}, \bibinfo {author}
  {\bibfnamefont {V.}~\bibnamefont {Kashcheyevs}}, \bibinfo {author}
  {\bibfnamefont {E.~V.}\ \bibnamefont {Kurganova}}, \bibinfo {author}
  {\bibfnamefont {U.}~\bibnamefont {Zeitler}}, \bibinfo {author} {\bibfnamefont
  {T.}~\bibnamefont {Weimann}}, \bibinfo {author} {\bibfnamefont
  {K.}~\bibnamefont {Pierz}},\ and\ \bibinfo {author} {\bibfnamefont {H.~W.}\
  \bibnamefont {Schumacher}},\ }\bibfield  {title} {\bibinfo {title}
  {{Generation of energy selective excitations in quantum Hall edge states}},\
  }\href {https://doi.org/10.1088/0268-1242/26/5/055010} {\bibfield  {journal}
  {\bibinfo  {journal} {Semiconductor Science and Technology}\ }\textbf
  {\bibinfo {volume} {26}},\ \bibinfo {pages} {055010} (\bibinfo {year}
  {2011})}\BibitemShut {NoStop}%
\bibitem [{\citenamefont {Fletcher}\ \emph {et~al.}(2013)\citenamefont
  {Fletcher}, \citenamefont {See}, \citenamefont {Howe}, \citenamefont
  {Pepper}, \citenamefont {Giblin}, \citenamefont {Griffiths}, \citenamefont
  {Jones}, \citenamefont {Farrer}, \citenamefont {Ritchie}, \citenamefont
  {Janssen},\ and\ \citenamefont {Kataoka}}]{Fletcher2012}%
  \BibitemOpen
  \bibfield  {author} {\bibinfo {author} {\bibfnamefont {J.~D.}\ \bibnamefont
  {Fletcher}}, \bibinfo {author} {\bibfnamefont {P.}~\bibnamefont {See}},
  \bibinfo {author} {\bibfnamefont {H.}~\bibnamefont {Howe}}, \bibinfo {author}
  {\bibfnamefont {M.}~\bibnamefont {Pepper}}, \bibinfo {author} {\bibfnamefont
  {S.~P.}\ \bibnamefont {Giblin}}, \bibinfo {author} {\bibfnamefont {J.~P.}\
  \bibnamefont {Griffiths}}, \bibinfo {author} {\bibfnamefont {G.~A.~C.}\
  \bibnamefont {Jones}}, \bibinfo {author} {\bibfnamefont {I.}~\bibnamefont
  {Farrer}}, \bibinfo {author} {\bibfnamefont {D.~A.}\ \bibnamefont {Ritchie}},
  \bibinfo {author} {\bibfnamefont {T.~J. B.~M.}\ \bibnamefont {Janssen}},\
  and\ \bibinfo {author} {\bibfnamefont {M.}~\bibnamefont {Kataoka}},\
  }\bibfield  {title} {\bibinfo {title} {{Clock-Controlled Emission of
  Single-Electron Wave Packets in a Solid-State Circuit}},\ }\href
  {https://doi.org/10.1103/PhysRevLett.111.216807} {\bibfield  {journal}
  {\bibinfo  {journal} {Physical Review Letters}\ }\textbf {\bibinfo {volume}
  {111}},\ \bibinfo {pages} {216807} (\bibinfo {year} {2013})}\BibitemShut
  {NoStop}%
\bibitem [{\citenamefont {Ubbelohde}\ \emph {et~al.}(2015)\citenamefont
  {Ubbelohde}, \citenamefont {Hohls}, \citenamefont {Kashcheyevs},
  \citenamefont {Wagner}, \citenamefont {Fricke}, \citenamefont
  {K{\"{a}}stner}, \citenamefont {Pierz}, \citenamefont {Schumacher},\ and\
  \citenamefont {Haug}}]{Ubbelohde2015}%
  \BibitemOpen
  \bibfield  {author} {\bibinfo {author} {\bibfnamefont {N.}~\bibnamefont
  {Ubbelohde}}, \bibinfo {author} {\bibfnamefont {F.}~\bibnamefont {Hohls}},
  \bibinfo {author} {\bibfnamefont {V.}~\bibnamefont {Kashcheyevs}}, \bibinfo
  {author} {\bibfnamefont {T.}~\bibnamefont {Wagner}}, \bibinfo {author}
  {\bibfnamefont {L.}~\bibnamefont {Fricke}}, \bibinfo {author} {\bibfnamefont
  {B.}~\bibnamefont {K{\"{a}}stner}}, \bibinfo {author} {\bibfnamefont
  {K.}~\bibnamefont {Pierz}}, \bibinfo {author} {\bibfnamefont {H.~W.}\
  \bibnamefont {Schumacher}},\ and\ \bibinfo {author} {\bibfnamefont {R.~J.}\
  \bibnamefont {Haug}},\ }\bibfield  {title} {\bibinfo {title} {{Partitioning
  of on-demand electron pairs}},\ }\href
  {https://doi.org/10.1038/nnano.2014.275} {\bibfield  {journal} {\bibinfo
  {journal} {Nature Nanotechnology}\ }\textbf {\bibinfo {volume} {10}},\
  \bibinfo {pages} {46} (\bibinfo {year} {2015})}\BibitemShut {NoStop}%
\bibitem [{\citenamefont {Waldie}\ \emph {et~al.}(2015)\citenamefont {Waldie},
  \citenamefont {See}, \citenamefont {Kashcheyevs}, \citenamefont {Griffiths},
  \citenamefont {Farrer}, \citenamefont {Jones}, \citenamefont {Ritchie},
  \citenamefont {Janssen},\ and\ \citenamefont {Kataoka}}]{Waldie2015}%
  \BibitemOpen
  \bibfield  {author} {\bibinfo {author} {\bibfnamefont {J.}~\bibnamefont
  {Waldie}}, \bibinfo {author} {\bibfnamefont {P.}~\bibnamefont {See}},
  \bibinfo {author} {\bibfnamefont {V.}~\bibnamefont {Kashcheyevs}}, \bibinfo
  {author} {\bibfnamefont {J.~P.}\ \bibnamefont {Griffiths}}, \bibinfo {author}
  {\bibfnamefont {I.}~\bibnamefont {Farrer}}, \bibinfo {author} {\bibfnamefont
  {G.~A.~C.}\ \bibnamefont {Jones}}, \bibinfo {author} {\bibfnamefont {D.~A.}\
  \bibnamefont {Ritchie}}, \bibinfo {author} {\bibfnamefont {T.~J. B.~M.}\
  \bibnamefont {Janssen}},\ and\ \bibinfo {author} {\bibfnamefont
  {M.}~\bibnamefont {Kataoka}},\ }\bibfield  {title} {\bibinfo {title}
  {{Measurement and control of electron wave packets from a single-electron
  source}},\ }\href {https://doi.org/10.1103/PhysRevB.92.125305} {\bibfield
  {journal} {\bibinfo  {journal} {Physical Review B}\ }\textbf {\bibinfo
  {volume} {92}},\ \bibinfo {pages} {125305} (\bibinfo {year}
  {2015})}\BibitemShut {NoStop}%
\bibitem [{\citenamefont {Kataoka}\ \emph {et~al.}(2016)\citenamefont
  {Kataoka}, \citenamefont {Johnson}, \citenamefont {Emary}, \citenamefont
  {See}, \citenamefont {Griffiths}, \citenamefont {Jones}, \citenamefont
  {Farrer}, \citenamefont {Ritchie}, \citenamefont {Pepper},\ and\
  \citenamefont {Janssen}}]{Kataoka2016a}%
  \BibitemOpen
  \bibfield  {author} {\bibinfo {author} {\bibfnamefont {M.}~\bibnamefont
  {Kataoka}}, \bibinfo {author} {\bibfnamefont {N.}~\bibnamefont {Johnson}},
  \bibinfo {author} {\bibfnamefont {C.}~\bibnamefont {Emary}}, \bibinfo
  {author} {\bibfnamefont {P.}~\bibnamefont {See}}, \bibinfo {author}
  {\bibfnamefont {J.~P.}\ \bibnamefont {Griffiths}}, \bibinfo {author}
  {\bibfnamefont {G.~A.~C.}\ \bibnamefont {Jones}}, \bibinfo {author}
  {\bibfnamefont {I.}~\bibnamefont {Farrer}}, \bibinfo {author} {\bibfnamefont
  {D.~A.}\ \bibnamefont {Ritchie}}, \bibinfo {author} {\bibfnamefont
  {M.}~\bibnamefont {Pepper}},\ and\ \bibinfo {author} {\bibfnamefont {T.~J.
  B.~M.}\ \bibnamefont {Janssen}},\ }\bibfield  {title} {\bibinfo {title}
  {{Time-of-Flight Measurements of Single-Electron Wave Packets in Quantum Hall
  Edge States}},\ }\href {https://doi.org/10.1103/PhysRevLett.116.126803}
  {\bibfield  {journal} {\bibinfo  {journal} {Physical Review Letters}\
  }\textbf {\bibinfo {volume} {116}},\ \bibinfo {pages} {126803} (\bibinfo
  {year} {2016})}\BibitemShut {NoStop}%
\bibitem [{\citenamefont {Freise}\ \emph {et~al.}(2020)\citenamefont {Freise},
  \citenamefont {Gerster}, \citenamefont {Reifert}, \citenamefont {Weimann},
  \citenamefont {Pierz}, \citenamefont {Hohls},\ and\ \citenamefont
  {Ubbelohde}}]{Freise2019}%
  \BibitemOpen
  \bibfield  {author} {\bibinfo {author} {\bibfnamefont {L.}~\bibnamefont
  {Freise}}, \bibinfo {author} {\bibfnamefont {T.}~\bibnamefont {Gerster}},
  \bibinfo {author} {\bibfnamefont {D.}~\bibnamefont {Reifert}}, \bibinfo
  {author} {\bibfnamefont {T.}~\bibnamefont {Weimann}}, \bibinfo {author}
  {\bibfnamefont {K.}~\bibnamefont {Pierz}}, \bibinfo {author} {\bibfnamefont
  {F.}~\bibnamefont {Hohls}},\ and\ \bibinfo {author} {\bibfnamefont
  {N.}~\bibnamefont {Ubbelohde}},\ }\bibfield  {title} {\bibinfo {title}
  {{Trapping and Counting Ballistic Nonequilibrium Electrons}},\ }\href
  {https://doi.org/10.1103/PhysRevLett.124.127701} {\bibfield  {journal}
  {\bibinfo  {journal} {Physical Review Letters}\ }\textbf {\bibinfo {volume}
  {124}},\ \bibinfo {pages} {127701} (\bibinfo {year} {2020})}\BibitemShut
  {NoStop}%
\bibitem [{\citenamefont {Kaestner}\ and\ \citenamefont
  {Kashcheyevs}(2015)}]{Kaestner2015}%
  \BibitemOpen
  \bibfield  {author} {\bibinfo {author} {\bibfnamefont {B.}~\bibnamefont
  {Kaestner}}\ and\ \bibinfo {author} {\bibfnamefont {V.}~\bibnamefont
  {Kashcheyevs}},\ }\bibfield  {title} {\bibinfo {title} {{Non-adiabatic
  quantized charge pumping with tunable-barrier quantum dots: a review of
  current progress}},\ }\href {https://doi.org/10.1088/0034-4885/78/10/103901}
  {\bibfield  {journal} {\bibinfo  {journal} {Reports on Progress in Physics}\
  }\textbf {\bibinfo {volume} {78}},\ \bibinfo {pages} {103901} (\bibinfo
  {year} {2015})}\BibitemShut {NoStop}%
\bibitem [{\citenamefont {Giblin}\ \emph {et~al.}(2019)\citenamefont {Giblin},
  \citenamefont {Fujiwara}, \citenamefont {Yamahata}, \citenamefont {Bae},
  \citenamefont {Kim}, \citenamefont {Rossi}, \citenamefont
  {M{\"{o}}tt{\"{o}}nen},\ and\ \citenamefont {Kataoka}}]{Giblin2019}%
  \BibitemOpen
  \bibfield  {author} {\bibinfo {author} {\bibfnamefont {S.~P.}\ \bibnamefont
  {Giblin}}, \bibinfo {author} {\bibfnamefont {A.}~\bibnamefont {Fujiwara}},
  \bibinfo {author} {\bibfnamefont {G.}~\bibnamefont {Yamahata}}, \bibinfo
  {author} {\bibfnamefont {M.-H.}\ \bibnamefont {Bae}}, \bibinfo {author}
  {\bibfnamefont {N.}~\bibnamefont {Kim}}, \bibinfo {author} {\bibfnamefont
  {A.}~\bibnamefont {Rossi}}, \bibinfo {author} {\bibfnamefont
  {M.}~\bibnamefont {M{\"{o}}tt{\"{o}}nen}},\ and\ \bibinfo {author}
  {\bibfnamefont {M.}~\bibnamefont {Kataoka}},\ }\bibfield  {title} {\bibinfo
  {title} {{Evidence for universality of tunable-barrier electron pumps}},\
  }\href {https://doi.org/10.1088/1681-7575/ab29a5} {\bibfield  {journal}
  {\bibinfo  {journal} {Metrologia}\ }\textbf {\bibinfo {volume} {56}},\
  \bibinfo {pages} {044004} (\bibinfo {year} {2019})}\BibitemShut {NoStop}%
\bibitem [{\citenamefont {Reifert}\ \emph {et~al.}(2021)\citenamefont
  {Reifert}, \citenamefont {Kokainis}, \citenamefont {Ambainis}, \citenamefont
  {Kashcheyevs},\ and\ \citenamefont {Ubbelohde}}]{Reifert2019}%
  \BibitemOpen
  \bibfield  {author} {\bibinfo {author} {\bibfnamefont {D.}~\bibnamefont
  {Reifert}}, \bibinfo {author} {\bibfnamefont {M.}~\bibnamefont {Kokainis}},
  \bibinfo {author} {\bibfnamefont {A.}~\bibnamefont {Ambainis}}, \bibinfo
  {author} {\bibfnamefont {V.}~\bibnamefont {Kashcheyevs}},\ and\ \bibinfo
  {author} {\bibfnamefont {N.}~\bibnamefont {Ubbelohde}},\ }\bibfield  {title}
  {\bibinfo {title} {{A random-walk benchmark for single-electron circuits}},\
  }\href {https://doi.org/10.1038/s41467-020-20554-w} {\bibfield  {journal}
  {\bibinfo  {journal} {Nature Communications}\ }\textbf {\bibinfo {volume}
  {12}},\ \bibinfo {pages} {285} (\bibinfo {year} {2021})}\BibitemShut
  {NoStop}%
\bibitem [{\citenamefont {Kataoka}\ \emph {et~al.}(2017)\citenamefont
  {Kataoka}, \citenamefont {Fletcher},\ and\ \citenamefont
  {Johnson}}]{Kataoka2016pss}%
  \BibitemOpen
  \bibfield  {author} {\bibinfo {author} {\bibfnamefont {M.}~\bibnamefont
  {Kataoka}}, \bibinfo {author} {\bibfnamefont {J.~D.}\ \bibnamefont
  {Fletcher}},\ and\ \bibinfo {author} {\bibfnamefont {N.}~\bibnamefont
  {Johnson}},\ }\bibfield  {title} {\bibinfo {title} {{Time-resolved
  single-electron wave-packet detection}},\ }\href
  {https://doi.org/10.1002/pssb.201600547} {\bibfield  {journal} {\bibinfo
  {journal} {physica status solidi (b)}\ }\textbf {\bibinfo {volume} {254}},\
  \bibinfo {pages} {1600547} (\bibinfo {year} {2017})}\BibitemShut {NoStop}%
\bibitem [{\citenamefont {Fletcher}\ \emph {et~al.}(2019)\citenamefont
  {Fletcher}, \citenamefont {Johnson}, \citenamefont {Locane}, \citenamefont
  {See}, \citenamefont {Griffiths}, \citenamefont {Farrer}, \citenamefont
  {Ritchie}, \citenamefont {Brouwer}, \citenamefont {Kashcheyevs},\ and\
  \citenamefont {Kataoka}}]{Fletcher2019}%
  \BibitemOpen
  \bibfield  {author} {\bibinfo {author} {\bibfnamefont {J.~D.}\ \bibnamefont
  {Fletcher}}, \bibinfo {author} {\bibfnamefont {N.}~\bibnamefont {Johnson}},
  \bibinfo {author} {\bibfnamefont {E.}~\bibnamefont {Locane}}, \bibinfo
  {author} {\bibfnamefont {P.}~\bibnamefont {See}}, \bibinfo {author}
  {\bibfnamefont {J.~P.}\ \bibnamefont {Griffiths}}, \bibinfo {author}
  {\bibfnamefont {I.}~\bibnamefont {Farrer}}, \bibinfo {author} {\bibfnamefont
  {D.~A.}\ \bibnamefont {Ritchie}}, \bibinfo {author} {\bibfnamefont {P.~W.}\
  \bibnamefont {Brouwer}}, \bibinfo {author} {\bibfnamefont {V.}~\bibnamefont
  {Kashcheyevs}},\ and\ \bibinfo {author} {\bibfnamefont {M.}~\bibnamefont
  {Kataoka}},\ }\bibfield  {title} {\bibinfo {title} {{Continuous-variable
  tomography of solitary electrons}},\ }\href
  {https://doi.org/10.1038/s41467-019-13222-1} {\bibfield  {journal} {\bibinfo
  {journal} {Nature Communications}\ }\textbf {\bibinfo {volume} {10}},\
  \bibinfo {pages} {5298} (\bibinfo {year} {2019})}\BibitemShut {NoStop}%
\bibitem [{\citenamefont {Locane}\ \emph {et~al.}(2019)\citenamefont {Locane},
  \citenamefont {Brouwer},\ and\ \citenamefont {Kashcheyevs}}]{Locane2019}%
  \BibitemOpen
  \bibfield  {author} {\bibinfo {author} {\bibfnamefont {E.}~\bibnamefont
  {Locane}}, \bibinfo {author} {\bibfnamefont {P.~W.}\ \bibnamefont
  {Brouwer}},\ and\ \bibinfo {author} {\bibfnamefont {V.}~\bibnamefont
  {Kashcheyevs}},\ }\bibfield  {title} {\bibinfo {title} {{Time-energy
  filtering of single electrons in ballistic waveguides}},\ }\href
  {https://doi.org/10.1088/1367-2630/ab3fbb} {\bibfield  {journal} {\bibinfo
  {journal} {New Journal of Physics}\ }\textbf {\bibinfo {volume} {21}},\
  \bibinfo {pages} {093042} (\bibinfo {year} {2019})}\BibitemShut {NoStop}%
\bibitem [{\citenamefont {Pavlovska}\ \emph {et~al.}(2023)\citenamefont
  {Pavlovska}, \citenamefont {Silvestrov}, \citenamefont {Recher},
  \citenamefont {Barinovs},\ and\ \citenamefont {Kashcheyevs}}]{Pavlovska2022}%
  \BibitemOpen
  \bibfield  {author} {\bibinfo {author} {\bibfnamefont {E.}~\bibnamefont
  {Pavlovska}}, \bibinfo {author} {\bibfnamefont {P.~G.}\ \bibnamefont
  {Silvestrov}}, \bibinfo {author} {\bibfnamefont {P.}~\bibnamefont {Recher}},
  \bibinfo {author} {\bibfnamefont {G.}~\bibnamefont {Barinovs}},\ and\
  \bibinfo {author} {\bibfnamefont {V.}~\bibnamefont {Kashcheyevs}},\
  }\bibfield  {title} {\bibinfo {title} {{Collision of two interacting
  electrons on a mesoscopic beam splitter: Exact solution in the classical
  limit}},\ }\href {https://doi.org/10.1103/PhysRevB.107.165304} {\bibfield
  {journal} {\bibinfo  {journal} {Physical Review B}\ }\textbf {\bibinfo
  {volume} {107}},\ \bibinfo {pages} {165304} (\bibinfo {year}
  {2023})}\BibitemShut {NoStop}%
\bibitem [{\citenamefont {Ubbelohde}\ \emph {et~al.}(2023)\citenamefont
  {Ubbelohde}, \citenamefont {Freise}, \citenamefont {Pavlovska}, \citenamefont
  {Silvestrov}, \citenamefont {Recher}, \citenamefont {Kokainis}, \citenamefont
  {Barinovs}, \citenamefont {Hohls}, \citenamefont {Weimann}, \citenamefont
  {Pierz},\ and\ \citenamefont {Kashcheyevs}}]{Ubbelohde2022}%
  \BibitemOpen
  \bibfield  {author} {\bibinfo {author} {\bibfnamefont {N.}~\bibnamefont
  {Ubbelohde}}, \bibinfo {author} {\bibfnamefont {L.}~\bibnamefont {Freise}},
  \bibinfo {author} {\bibfnamefont {E.}~\bibnamefont {Pavlovska}}, \bibinfo
  {author} {\bibfnamefont {P.~G.}\ \bibnamefont {Silvestrov}}, \bibinfo
  {author} {\bibfnamefont {P.}~\bibnamefont {Recher}}, \bibinfo {author}
  {\bibfnamefont {M.}~\bibnamefont {Kokainis}}, \bibinfo {author}
  {\bibfnamefont {G.}~\bibnamefont {Barinovs}}, \bibinfo {author}
  {\bibfnamefont {F.}~\bibnamefont {Hohls}}, \bibinfo {author} {\bibfnamefont
  {T.}~\bibnamefont {Weimann}}, \bibinfo {author} {\bibfnamefont
  {K.}~\bibnamefont {Pierz}},\ and\ \bibinfo {author} {\bibfnamefont
  {V.}~\bibnamefont {Kashcheyevs}},\ }\bibfield  {title} {\bibinfo {title}
  {{Two electrons interacting at a mesoscopic beam splitter}},\ }\href
  {https://doi.org/10.1038/s41565-023-01370-x} {\bibfield  {journal} {\bibinfo
  {journal} {Nature Nanotechnology}\ }\textbf {\bibinfo {volume} {18}},\
  \bibinfo {pages} {733} (\bibinfo {year} {2023})}\BibitemShut {NoStop}%
\bibitem [{\citenamefont {Fletcher}\ \emph {et~al.}(2023)\citenamefont
  {Fletcher}, \citenamefont {Park}, \citenamefont {Ryu}, \citenamefont {See},
  \citenamefont {Griffiths}, \citenamefont {Jones}, \citenamefont {Farrer},
  \citenamefont {Ritchie}, \citenamefont {Sim},\ and\ \citenamefont
  {Kataoka}}]{Fletcher2022}%
  \BibitemOpen
  \bibfield  {author} {\bibinfo {author} {\bibfnamefont {J.~D.}\ \bibnamefont
  {Fletcher}}, \bibinfo {author} {\bibfnamefont {W.}~\bibnamefont {Park}},
  \bibinfo {author} {\bibfnamefont {S.}~\bibnamefont {Ryu}}, \bibinfo {author}
  {\bibfnamefont {P.}~\bibnamefont {See}}, \bibinfo {author} {\bibfnamefont
  {J.~P.}\ \bibnamefont {Griffiths}}, \bibinfo {author} {\bibfnamefont
  {G.~A.~C.}\ \bibnamefont {Jones}}, \bibinfo {author} {\bibfnamefont
  {I.}~\bibnamefont {Farrer}}, \bibinfo {author} {\bibfnamefont {D.~A.}\
  \bibnamefont {Ritchie}}, \bibinfo {author} {\bibfnamefont {H.-S.}\
  \bibnamefont {Sim}},\ and\ \bibinfo {author} {\bibfnamefont {M.}~\bibnamefont
  {Kataoka}},\ }\bibfield  {title} {\bibinfo {title} {{Time-resolved Coulomb
  collision of single electrons}},\ }\href
  {https://doi.org/10.1038/s41565-023-01369-4} {\bibfield  {journal} {\bibinfo
  {journal} {Nature Nanotechnology}\ }\textbf {\bibinfo {volume} {18}},\
  \bibinfo {pages} {727} (\bibinfo {year} {2023})}\BibitemShut {NoStop}%
\bibitem [{\citenamefont {Fletcher}\ \emph {et~al.}(2024)\citenamefont
  {Fletcher}, \citenamefont {Park}, \citenamefont {See}, \citenamefont
  {Griffiths}, \citenamefont {Jones}, \citenamefont {Farrer}, \citenamefont
  {Ritchie}, \citenamefont {Sim},\ and\ \citenamefont
  {Kataoka}}]{Fletcher2024}%
  \BibitemOpen
  \bibfield  {author} {\bibinfo {author} {\bibfnamefont {J.~D.}\ \bibnamefont
  {Fletcher}}, \bibinfo {author} {\bibfnamefont {W.}~\bibnamefont {Park}},
  \bibinfo {author} {\bibfnamefont {P.}~\bibnamefont {See}}, \bibinfo {author}
  {\bibfnamefont {J.~P.}\ \bibnamefont {Griffiths}}, \bibinfo {author}
  {\bibfnamefont {G.~A.~C.}\ \bibnamefont {Jones}}, \bibinfo {author}
  {\bibfnamefont {I.}~\bibnamefont {Farrer}}, \bibinfo {author} {\bibfnamefont
  {D.~A.}\ \bibnamefont {Ritchie}}, \bibinfo {author} {\bibfnamefont {H.~S.}\
  \bibnamefont {Sim}},\ and\ \bibinfo {author} {\bibfnamefont {M.}~\bibnamefont
  {Kataoka}},\ }\bibfield  {title} {\bibinfo {title} {Coulomb sensing of single
  ballistic electrons},\ }\Eprint {https://arxiv.org/abs/2412.15789}
  {arXiv:2412.15789}  (\bibinfo {year} {2024})\BibitemShut {NoStop}%
\bibitem [{\citenamefont {Laughlin}(1983{\natexlab{a}})}]{Laughlin1983b}%
  \BibitemOpen
  \bibfield  {author} {\bibinfo {author} {\bibfnamefont {R.~B.}\ \bibnamefont
  {Laughlin}},\ }\bibfield  {title} {\bibinfo {title} {{Quantized motion of
  three two-dimensional electrons in a strong magnetic field}},\ }\href
  {https://doi.org/10.1103/PhysRevB.27.3383} {\bibfield  {journal} {\bibinfo
  {journal} {Physical Review B}\ }\textbf {\bibinfo {volume} {27}},\ \bibinfo
  {pages} {3383} (\bibinfo {year} {1983}{\natexlab{a}})}\BibitemShut {NoStop}%
\bibitem [{\citenamefont {Laughlin}(1983{\natexlab{b}})}]{Laughlin1983a}%
  \BibitemOpen
  \bibfield  {author} {\bibinfo {author} {\bibfnamefont {R.~B.}\ \bibnamefont
  {Laughlin}},\ }\bibfield  {title} {\bibinfo {title} {{Anomalous Quantum Hall
  Effect: An Incompressible Quantum Fluid with Fractionally Charged
  Excitations}},\ }\href {https://doi.org/10.1103/PhysRevLett.50.1395}
  {\bibfield  {journal} {\bibinfo  {journal} {Physical Review Letters}\
  }\textbf {\bibinfo {volume} {50}},\ \bibinfo {pages} {1395} (\bibinfo {year}
  {1983}{\natexlab{b}})}\BibitemShut {NoStop}%
\bibitem [{\citenamefont {Fertig}\ and\ \citenamefont
  {Halperin}(1987)}]{Fertig1987}%
  \BibitemOpen
  \bibfield  {author} {\bibinfo {author} {\bibfnamefont {H.~A.}\ \bibnamefont
  {Fertig}}\ and\ \bibinfo {author} {\bibfnamefont {B.~I.}\ \bibnamefont
  {Halperin}},\ }\bibfield  {title} {\bibinfo {title} {{Transmission
  coefficient of an electron through a saddle-point potential in a magnetic
  field}},\ }\href {https://doi.org/10.1103/PhysRevB.36.7969} {\bibfield
  {journal} {\bibinfo  {journal} {Physical Review B}\ }\textbf {\bibinfo
  {volume} {36}},\ \bibinfo {pages} {7969} (\bibinfo {year}
  {1987})}\BibitemShut {NoStop}%
\bibitem [{Note1()}]{Note1}%
  \BibitemOpen
  \bibinfo {note} {To characterize the curvature of the potential both
  experimental~\cite {Kataoka2016a} and theoretical~\cite {Fertig1987} papers
  often introduce the harmonic-oscillator-like frequency. That means Eq.~(\ref
  {saddle_point}) became $ V(x,y)=\protect \frac {m}{2}(\omega _y^2 y^2 -\omega
  _x^2 x^2)$ with obvious translation rules $A=m\omega _x^2$, $C=m\omega
  _y^2$.}\BibitemShut {Stop}%
\bibitem [{\citenamefont {Huckestein}(1995)}]{Huckestein1995}%
  \BibitemOpen
  \bibfield  {author} {\bibinfo {author} {\bibfnamefont {B.}~\bibnamefont
  {Huckestein}},\ }\bibfield  {title} {\bibinfo {title} {{Scaling theory of the
  integer quantum Hall effect}},\ }\href
  {https://doi.org/10.1103/RevModPhys.67.357} {\bibfield  {journal} {\bibinfo
  {journal} {Reviews of Modern Physics}\ }\textbf {\bibinfo {volume} {67}},\
  \bibinfo {pages} {357} (\bibinfo {year} {1995})}\BibitemShut {NoStop}%
\bibitem [{\citenamefont {Emary}\ \emph {et~al.}(2019)\citenamefont {Emary},
  \citenamefont {Clark}, \citenamefont {Kataoka},\ and\ \citenamefont
  {Johnson}}]{Emary2019}%
  \BibitemOpen
  \bibfield  {author} {\bibinfo {author} {\bibfnamefont {C.}~\bibnamefont
  {Emary}}, \bibinfo {author} {\bibfnamefont {L.~A.}\ \bibnamefont {Clark}},
  \bibinfo {author} {\bibfnamefont {M.}~\bibnamefont {Kataoka}},\ and\ \bibinfo
  {author} {\bibfnamefont {N.}~\bibnamefont {Johnson}},\ }\bibfield  {title}
  {\bibinfo {title} {{Energy relaxation in hot electron quantum optics via
  acoustic and optical phonon emission}},\ }\href
  {https://doi.org/10.1103/PhysRevB.99.045306} {\bibfield  {journal} {\bibinfo
  {journal} {Physical Review B}\ }\textbf {\bibinfo {volume} {99}},\ \bibinfo
  {pages} {045306} (\bibinfo {year} {2019})}\BibitemShut {NoStop}%
\bibitem [{\citenamefont {Hillery}\ \emph {et~al.}(1984)\citenamefont
  {Hillery}, \citenamefont {O'Connell}, \citenamefont {Scully},\ and\
  \citenamefont {Wigner}}]{Hillery1984}%
  \BibitemOpen
  \bibfield  {author} {\bibinfo {author} {\bibfnamefont {M.}~\bibnamefont
  {Hillery}}, \bibinfo {author} {\bibfnamefont {R.}~\bibnamefont {O'Connell}},
  \bibinfo {author} {\bibfnamefont {M.}~\bibnamefont {Scully}},\ and\ \bibinfo
  {author} {\bibfnamefont {E.}~\bibnamefont {Wigner}},\ }\bibfield  {title}
  {\bibinfo {title} {{Distribution functions in physics: Fundamentals}},\
  }\href {https://doi.org/10.1016/0370-1573(84)90160-1} {\bibfield  {journal}
  {\bibinfo  {journal} {Physics Reports}\ }\textbf {\bibinfo {volume} {106}},\
  \bibinfo {pages} {121} (\bibinfo {year} {1984})}\BibitemShut {NoStop}%
\bibitem [{\citenamefont {Husimi}(1953)}]{Husimi1953}%
  \BibitemOpen
  \bibfield  {author} {\bibinfo {author} {\bibfnamefont {K.}~\bibnamefont
  {Husimi}},\ }\bibfield  {title} {\bibinfo {title} {{Miscellanea in Elementary
  Quantum Mechanics, II}},\ }\href {https://doi.org/10.1143/ptp/9.4.381}
  {\bibfield  {journal} {\bibinfo  {journal} {Progress of Theoretical Physics}\
  }\textbf {\bibinfo {volume} {9}},\ \bibinfo {pages} {381} (\bibinfo {year}
  {1953})}\BibitemShut {NoStop}%
\bibitem [{\citenamefont {Lee}(1995)}]{Moyal}%
  \BibitemOpen
  \bibfield  {author} {\bibinfo {author} {\bibfnamefont {H.-W.}\ \bibnamefont
  {Lee}},\ }\bibfield  {title} {\bibinfo {title} {{Theory and application of
  the quantum phase-space distribution functions}},\ }\href
  {https://doi.org/10.1016/0370-1573(95)00007-4} {\bibfield  {journal}
  {\bibinfo  {journal} {Physics Reports}\ }\textbf {\bibinfo {volume} {259}},\
  \bibinfo {pages} {147} (\bibinfo {year} {1995})}\BibitemShut {NoStop}%
\bibitem [{\citenamefont {Heim}\ \emph {et~al.}(2013)\citenamefont {Heim},
  \citenamefont {Schleich}, \citenamefont {Alsing}, \citenamefont {Dahl},\ and\
  \citenamefont {Varro}}]{Heim2013}%
  \BibitemOpen
  \bibfield  {author} {\bibinfo {author} {\bibfnamefont {D.}~\bibnamefont
  {Heim}}, \bibinfo {author} {\bibfnamefont {W.}~\bibnamefont {Schleich}},
  \bibinfo {author} {\bibfnamefont {P.}~\bibnamefont {Alsing}}, \bibinfo
  {author} {\bibfnamefont {J.}~\bibnamefont {Dahl}},\ and\ \bibinfo {author}
  {\bibfnamefont {S.}~\bibnamefont {Varro}},\ }\bibfield  {title} {\bibinfo
  {title} {{Tunneling of an energy eigenstate through a parabolic barrier
  viewed from Wigner phase space}},\ }\href
  {https://doi.org/10.1016/j.physleta.2013.05.017} {\bibfield  {journal}
  {\bibinfo  {journal} {Physics Letters A}\ }\textbf {\bibinfo {volume}
  {377}},\ \bibinfo {pages} {1822} (\bibinfo {year} {2013})}\BibitemShut
  {NoStop}%
\bibitem [{Note2()}]{Note2}%
  \BibitemOpen
  \bibinfo {note} {In terms of the dimensionless function used in
  Refs.~\protect \rev@citealp {Pavlovska2022,Ubbelohde2022}, one has $\Phi _k=
  E_{+}/(2 \gamma _0 \varepsilon _0)$ for $|\Phi _k| \ll 1$ with
  $E_{+}=\varepsilon [1+\cosh (\lambda \Delta t_{\protect \text {in}})]$ if
  $\varepsilon =\varepsilon _{1\protect \text {in}}=\varepsilon _{2\protect
  \text {in}}$.}\BibitemShut {Stop}%
\bibitem [{Note3()}]{Note3}%
  \BibitemOpen
  \bibinfo {note} {Note that a change in energy leads to a logarithmic time
  shift even for one non-interacting electron as this is a basic dispersion
  property of an energy-selective beamsplitter (see Section~\ref {section5}
  above and Sec.~III-C of Ref.~\protect \rev@citealp
  {Pavlovska2022}).}\BibitemShut {Stop}%
\bibitem [{\citenamefont {Recher}\ \emph {et~al.}(2000)\citenamefont {Recher},
  \citenamefont {Sukhorukov},\ and\ \citenamefont {Loss}}]{Recher2000}%
  \BibitemOpen
  \bibfield  {author} {\bibinfo {author} {\bibfnamefont {P.}~\bibnamefont
  {Recher}}, \bibinfo {author} {\bibfnamefont {E.~V.}\ \bibnamefont
  {Sukhorukov}},\ and\ \bibinfo {author} {\bibfnamefont {D.}~\bibnamefont
  {Loss}},\ }\bibfield  {title} {\bibinfo {title} {Quantum dot as spin filter
  and spin memory},\ }\href {https://doi.org/10.1103/PhysRevLett.85.1962}
  {\bibfield  {journal} {\bibinfo  {journal} {Physical Review Letters}\ }\textbf
  {\bibinfo {volume} {85}},\ \bibinfo {pages} {1962} (\bibinfo {year}
  {2000})}\BibitemShut {NoStop}%
\bibitem [{\citenamefont {Wenz}\ \emph {et~al.}(2019)\citenamefont {Wenz},
  \citenamefont {Klochan}, \citenamefont {Hohls}, \citenamefont {Gerster},
  \citenamefont {Kashcheyevs},\ and\ \citenamefont {Schumacher}}]{Wenz2019}%
  \BibitemOpen
  \bibfield  {author} {\bibinfo {author} {\bibfnamefont {T.}~\bibnamefont
  {Wenz}}, \bibinfo {author} {\bibfnamefont {J.}~\bibnamefont {Klochan}},
  \bibinfo {author} {\bibfnamefont {F.}~\bibnamefont {Hohls}}, \bibinfo
  {author} {\bibfnamefont {T.}~\bibnamefont {Gerster}}, \bibinfo {author}
  {\bibfnamefont {V.}~\bibnamefont {Kashcheyevs}},\ and\ \bibinfo {author}
  {\bibfnamefont {H.~W.}\ \bibnamefont {Schumacher}},\ }\bibfield  {title}
  {\bibinfo {title} {{Quantum dot state initialization by control of tunneling
  rates}},\ }\href {https://doi.org/10.1103/PhysRevB.99.201409} {\bibfield
  {journal} {\bibinfo  {journal} {Physical Review B}\ }\textbf {\bibinfo
  {volume} {99}},\ \bibinfo {pages} {201409} (\bibinfo {year}
  {2019})}\BibitemShut {NoStop}%
\bibitem [{\citenamefont {Jadot}\ \emph {et~al.}(2021)\citenamefont {Jadot},
  \citenamefont {Mortemousque}, \citenamefont {Chanrion}, \citenamefont
  {Thiney}, \citenamefont {Ludwig}, \citenamefont {Wieck}, \citenamefont
  {Urdampilleta}, \citenamefont {B{\"{a}}uerle},\ and\ \citenamefont
  {Meunier}}]{Jadot2021}%
  \BibitemOpen
  \bibfield  {author} {\bibinfo {author} {\bibfnamefont {B.}~\bibnamefont
  {Jadot}}, \bibinfo {author} {\bibfnamefont {P.-A.}\ \bibnamefont
  {Mortemousque}}, \bibinfo {author} {\bibfnamefont {E.}~\bibnamefont
  {Chanrion}}, \bibinfo {author} {\bibfnamefont {V.}~\bibnamefont {Thiney}},
  \bibinfo {author} {\bibfnamefont {A.}~\bibnamefont {Ludwig}}, \bibinfo
  {author} {\bibfnamefont {A.~D.}\ \bibnamefont {Wieck}}, \bibinfo {author}
  {\bibfnamefont {M.}~\bibnamefont {Urdampilleta}}, \bibinfo {author}
  {\bibfnamefont {C.}~\bibnamefont {B{\"{a}}uerle}},\ and\ \bibinfo {author}
  {\bibfnamefont {T.}~\bibnamefont {Meunier}},\ }\bibfield  {title} {\bibinfo
  {title} {{Distant spin entanglement via fast and coherent electron
  shuttling}},\ }\href {https://doi.org/10.1038/s41565-021-00846-y} {\bibfield
  {journal} {\bibinfo  {journal} {Nature Nanotechnology}\ }\textbf {\bibinfo
  {volume} {16}},\ \bibinfo {pages} {570} (\bibinfo {year} {2021})}\BibitemShut
  {NoStop}%
\bibitem [{\citenamefont {Kawabata}(2001)}]{Kawabata2001}%
  \BibitemOpen
  \bibfield  {author} {\bibinfo {author} {\bibfnamefont {S.}~\bibnamefont
  {Kawabata}},\ }\bibfield  {title} {\bibinfo {title} {Test of bell's
  inequality using the spin filter effect in ferromagnetic semiconductor
  microstructures},\ }\href {https://doi.org/10.1143/JPSJ.70.1210} {\bibfield
  {journal} {\bibinfo  {journal} {Journal of the Physical Society of Japan}\
  }\textbf {\bibinfo {volume} {70}},\ \bibinfo {pages} {1210} (\bibinfo {year}
  {2001})}\BibitemShut {NoStop}%
\bibitem [{\citenamefont {Chtchelkatchev}\ \emph {et~al.}(2002)\citenamefont
  {Chtchelkatchev}, \citenamefont {Blatter}, \citenamefont {Lesovik},\ and\
  \citenamefont {Martin}}]{Chtchelkatchev2002}%
  \BibitemOpen
  \bibfield  {author} {\bibinfo {author} {\bibfnamefont {N.~M.}\ \bibnamefont
  {Chtchelkatchev}}, \bibinfo {author} {\bibfnamefont {G.}~\bibnamefont
  {Blatter}}, \bibinfo {author} {\bibfnamefont {G.~B.}\ \bibnamefont
  {Lesovik}},\ and\ \bibinfo {author} {\bibfnamefont {T.}~\bibnamefont
  {Martin}},\ }\bibfield  {title} {\bibinfo {title} {Bell inequalities and
  entanglement in solid-state devices},\ }\href
  {https://doi.org/10.1103/PhysRevB.66.161320} {\bibfield  {journal} {\bibinfo
  {journal} {Physical Review B}\ }\textbf {\bibinfo {volume} {66}},\ \bibinfo
  {pages} {161320} (\bibinfo {year} {2002})}\BibitemShut {NoStop}%
\bibitem [{\citenamefont {Recher}\ \emph {et~al.}(2004)\citenamefont {Recher},
  \citenamefont {Saraga},\ and\ \citenamefont {Loss}}]{Recher2004}%
  \BibitemOpen
  \bibfield  {author} {\bibinfo {author} {\bibfnamefont {P.}~\bibnamefont
  {Recher}}, \bibinfo {author} {\bibfnamefont {D.~S.}\ \bibnamefont {Saraga}},\
  and\ \bibinfo {author} {\bibfnamefont {D.}~\bibnamefont {Loss}},\ }\bibinfo
  {title} {Creation and detection of mobile and non-local spin-entangled
  electrons},\ in\ \href {https://doi.org/10.1007/1-4020-2193-3_11} {\emph
  {\bibinfo {booktitle} {Fundamental Problems of Mesoscopic Physics:
  Interactions and Decoherence}}},\ \bibinfo {editor} {edited by\ \bibinfo
  {editor} {\bibfnamefont {I.~V.}\ \bibnamefont {Lerner}}, \bibinfo {editor}
  {\bibfnamefont {B.~L.}\ \bibnamefont {Altshuler}},\ and\ \bibinfo {editor}
  {\bibfnamefont {Y.}~\bibnamefont {Gefen}}}\ (\bibinfo  {publisher} {Springer
  Netherlands},\ \bibinfo {address} {Dordrecht},\ \bibinfo {year} {2004})\ pp.\
  \bibinfo {pages} {179--202}\BibitemShut {NoStop}%
\end{thebibliography}

%

\end{document}